\documentclass[a4paper,11pt]{article}
\pdfoutput=1 

\usepackage{jcappub} 

\usepackage[T1]{fontenc} 

\usepackage{graphicx}
\usepackage{amssymb}
\usepackage{ulem}
\usepackage{multirow}
\usepackage{color}
\usepackage{rotating}
\usepackage{placeins} 

\usepackage{subfig}

\def\be{\begin{equation}}
\def\ee{\end{equation}}

\def\bi{\begin{itemize}}
\def\ei{\end{itemize}}

\def\ben{\begin{enumerate}}
\def\een{\end{enumerate}}

\def\bt{\begin{tabular}}
\def\et{\end{tabular}}

\def\bc{\begin{center}}
\def\ec{\end{center}}

\def\la{\label}

\def\bea{\begin{eqnarray}}
\def\eea{\end{eqnarray}}

\def\la{\langle}
\def\ra{\rangle}

\title{Non-Gaussian Shape Recognition}
\author{Joyce Byun and Rachel Bean}
\affiliation{Department of Astronomy, Cornell University, Ithaca, NY 14853, USA.}

\emailAdd{byun@astro.cornell.edu}
\emailAdd{rbean@astro.cornell.edu}

\abstract{A detection of primordial non-Gaussianity could transform our understanding of the fundamental theory of inflation. The precision promised by upcoming cosmic microwave background (CMB) and large-scale structure (LSS) surveys raises a natural question: if a detection given a particular template is made, what does this truly tell us about the underlying theory? Even in the case of non-detections and upper bounds on deviations from Gaussianity, what can we then infer about the viable theories that remain?
In this paper we present a systematic way to constrain a wide range of non-Gaussian shapes, including general single and multi-field models and models with excited initial states. We present a separable, divergent basis able to recreate many shapes in the literature to high accuracy with between three and seven basis functions. The basis allows shapes to be grouped into broad ``template classes'', satisfying theoretically-relevant priors on their divergence properties in the squeezed limit.
We forecast how well a Planck-like CMB survey could not only detect a general non-Gaussian signal but discern more about its shape, using existing templates and new ones we propose. This approach offers an opportunity to tie together minimal theoretical priors with observational constraints on the shape in general, and in the squeezed limit, to gain a deeper insight into what drove inflation.
}

\begin{document}
\maketitle
\flushbottom

\section{Introduction}
\label{sec:intro}

An early period of accelerated expansion, perhaps a trillionth of a second after the Big Bang, is proposed to solve a number of problems unresolved by the Big Bang scenario, such as the flatness and horizon problems. The paradigm of single-field slow-roll inflation is the simplest model to describe this acceleration,  and makes broad predictions of adiabatic and Gaussian-distributed primordial density (scalar) perturbations, described by a nearly scale-invariant 2-point correlation, and smaller gravitational metric (tensor) perturbations. In this model, the scalar and tensor amplitudes and scale dependence are related through a `consistency relationship'. 

The precision of astrophysical measurements has dramatically improved over the last decade.  Cosmic Microwave Background (CMB) measurements, such as from the Wilkinson Microwave Anisotropy Probe \cite{Bennett:2012fp,Hinshaw:2012fq},  small-scale CMB measurements from the Atacama Cosmology Telescope and South Pole Telescope \cite{Story:2012wx,Sievers:2013wk}, and large-scale structure (LSS) observations such as from the Sloan Digital Sky Survey \cite{Sanchez:2012sg,Ross:2012sx}, are all entirely consistent with single-field slow-roll predictions, placing strong constraints on the scalar power spectrum and upper limits on the degree of deviations from Gaussianity and amplitude of tensor modes.

The agreement between single-field inflationary predictions and observations is a profound success for cosmology, but it is as yet, insufficient to inform us about the underlying theory from which inflation derives. Rapid theoretical progress in high energy effective field theory has led to a wide range of possible Lagrangians for inflation \cite{Cheung:2007st,Weinberg:2008hq,Senatore:2010wk,Bartolo:2010bj}. These often go beyond making distinct predictions for the form of the single-field inflationary potential and can include multiple dynamical fields and non-canonical derivative (kinetic) terms in the action.

These alternative mechanisms can generate new observational signatures, including different consistency relationships relating scalar and tensor perturbations \cite{Garriga:1999vw,Bean:2007hc}, the addition of non-adiabatic (isocurvature) modes \cite{Polarski:1994rz,GarciaBellido:1995qq}, and the possibility of observationally measurable non-Gaussian correlations \cite{Komatsu:2001rj}. In particular, the possibility that primordial non-Gaussianity may be detectable as a non-zero 3-point correlation function, or bispectrum, has been a major development in the search for  observational signatures of the underlying inflationary theory. What is most exciting is that different theories can give rise to bispectra with distinct scale dependencies, such that measuring not only the amplitude but also the scale-dependence, or `shape', of the bispectrum could provide a direct insight into the inflationary mechanism.

Much work has focused on potentially measuring the amplitude of commonly predicted shapes, such as the local \cite{Gangui:1993tt,Verde:1999ij,Komatsu:2001rj}, equilateral  \cite{Creminelli:2005hu}, and orthogonal \cite{Senatore:2009gt} templates. Recent theoretical developments have also led to a wider population of bispectra, including those from fast-roll inflation \cite{Chen:2006nt,Khoury:2008wj,Noller:2011hd,Ribeiro:2012ar,Noller:2012ed}, quasi-single field inflation \cite{Chen:2009we,Chen:2009zp}, warm inflation \cite{Gupta:2002kn,Moss:2007cv,LopezNacir:2011kk}, and non-Bunch-Davies or excited initial states \cite{Chen:2006nt,Holman:2007na,Meerburg:2009ys,Agarwal:2012mq}. There are also hybrids of multi-field and non-slow-roll models \cite{Langlois:2008qf,Arroja:2008yy,RenauxPetel:2009sj}, and the inclusion of isocurvature modes in the non-Gaussian correlations \cite{Langlois:2011zz,Langlois:2011hn,Langlois:2012tm}. These bispectra can have very different shapes, meaning their signal is weighted towards different configurations of the 3 wavenumbers in (Fourier) $k$-space. How divergent shapes are in the `squeezed' $k$-configuration, when one of the three length scales contributing to the 3-point function becomes much larger than the other two, in particular can signal whether inflation is derived from a single-field or multi-field model.
  
The divergence in the squeezed limit could also be constrained by its effect on large-scale structure. A non-Gaussian signal peaking in the squeezed limit would directly couple large scale modes to small scales, on which non-linear halos are forming \cite{Dalal:2007cu}. This gives rise to an additional contribution to the halo bias, determining how the number density of halos of a given mass are related to the underlying linear power spectrum. In theory, wide field large-scale structure surveys could provide a sensitive constraint on the divergence properties of non-Gaussianity \cite{LoVerde:2007ri,Verde:2009hy,Becker:2010hx,Becker:2012yr,Norena:2012yi,Wagner:2011wx,Agullo:2012cs}.

Given the diversity of theoretically motivated shapes, an intriguing question is how well one might actually be able to determine the shape of primordial non-Gaussianity, rather than purely assuming a shape template is the true shape a priori. To what  extent can the shape of non-Gaussianity be reconstructed using the CMB and LSS 3-point correlations? If a positive detection is made assuming a template, how well would such a detection really constrain the underlying shape and the theoretical model that generated it?

This `reconstruction' approach has been widely considered in the context of the inflationary power spectrum, both in terms of $P(k)$ reconstruction (e.g. \cite{Spergel:2006hy,Hlozek:2011pc,Bridle:2003sa}), and measuring the hierarchy of slow-roll parameters (e.g. \cite{Copeland:1993zn,Kinney:2003uw,Bean:2008ga,Adshead:2008vn} and references therein), instead of assuming a nearly scale-invariant spectrum parametrized by a constant tilt $n_s$ and a constant running $dn_s/d\ln k$. 

Unfortunately, calculating theoretical predictions for CMB bispectra is computationally cumbersome in its exact form, requiring 4-dimensional integrals to be performed. A formalism to make the calculation tractable for general bispectra was introduced in \cite{Fergusson:2009nv}.  The authors proposed a technique to create templates for shapes by expanding non-separable shapes on a basis set of bispectra that are explicitly separable functions of the three wavenumbers. The separability reduces the 4-dimensional integral to a tractable computation without a significant reduction in the accuracy of the computed CMB bispectra. This approach has been used to forecast bispectrum constraints for a variety of fundamental shapes \cite{Hinshaw:2012fq} and adapted to other basis sets to describe oscillatory, rather than monotonic, shapes \cite{Meerburg:2010ks}.  Furthermore, the method of modal expansions on a separable basis has been shown to be advantageous and applicable in a variety of contexts, for example in studying CMB 3-point correlations with wavelets \cite{Regan:2013wwa}, CMB trispectra \cite{Regan:2010cn,Fergusson:2010gn}, and matter density bispectra in LSS \cite{Fergusson:2010ia,Regan:2011zq,Schmittfull:2012hq}.

In this work we present an alternative separable basis to efficiently describe and investigate the broad class of nearly scale-invariant general bispectra in terms of their squeezed limit properties. We discuss a way to expand a general shape in the basis, which is specifically tuned to enable us to systematically increase the complexity of the template in a theoretically motivated way. We forecast the potential for determining the underlying non-Gaussian shape given upcoming CMB temperature and E-mode polarization data modeled on the Planck survey.   

The format of the paper is as follows. In section \ref{formalism}, we review  the formalism used to calculate CMB bispectra.  We introduce a separable basis to describe general shapes that are scale-invariant and potentially divergent, and discuss how this basis can be applied to describe a wide variety of shapes in the literature. Using the basis, we develop an expansion that allows us to incrementally investigate classes of bispectra motivated by theories. In section \ref{analysis}, we present a Fisher analysis quantifying how well a Planck-like survey will be able to distinguish between and constrain individual bispectrum shapes.  Using a principal component analysis, we find the best to worst measured uncorrelated shapes, and compute the overall uncertainties in the bispectrum measurement as a function of $k$-space configuration under different theoretical priors. We use these results to establish how much we can learn about the bispectrum shape, and hence with what confidence we might be able to narrow down the underlying inflationary theory. In section \ref{conclusion}, we summarize our findings and discuss implications for future work.

\section{Efficient calculation of a general non-Gaussian shape}
\label{formalism}

In this section we lay out the formalism to describe and compute general bispectra. In subsections \ref{sec:bispec} and \ref{sec:similarity}, we respectively review the calculation of the CMB bispectrum given the primordial 3-point function and the definitions of covariances in wavenumber and multipole space that roughly quantify the theoretical similarity of two bispectra. In subsection \ref{sec:basis} we introduce a separable basis set to describe general bispectra and develop computationally tractable templates. Subsection \ref{sec:motivation} discusses the application of the basis set to a variety of theoretical bispectra and templates in the literature. How bispectra can be classified and presented pictorially is reviewed in subsection \ref{sec:classification}. 

\subsection{The CMB bispectrum}
\label{sec:bispec} 

While Gaussian fluctuations are wholly described by a 2-point correlation function, a full description of non-Gaussian fluctuations requires higher order correlations that are not trivially related to the 2-point function. The simplest higher order correlation is the 3-point function, where the 3-point Fourier space statistic analogous to the 2-point power spectrum is the bispectrum, $B_\Phi$, defined by
\be \label{eq:defBphi}
\la \Phi({\bf k}_1) \Phi({\bf k}_2) \Phi({\bf k}_3)\ra \equiv (2\pi)^3 \delta({\bf k_1+k_2+k_3}) B_{\Phi}(k_1,k_2,k_3).
\ee
$\Phi({\bf k})$ is the primordial gravitational potential, related to the curvature perturbation by $\Phi=\frac{3}{5}{\cal R}$. Under the assumptions of statistical isotropy and homogeneity, the bispectrum is dependent on only the magnitudes of the wavenumbers, $k_1$, $k_2$, and $k_3$.
 
The bispectrum is often parameterized by a shape, $S(k_1,k_2,k_3)$,  and an amplitude, $f_{NL}$, at an arbitrary configuration in $k$-space which together  determine the bispectrum at all scales, 
\bea
\frac{(k_1 k_2 k_3)^2} {N}B_{\Phi}(k_1,k_2,k_3) &=&f_{NL} S(k_1,k_2,k_3).\label{eq:BPhi}
\eea
The typical convention is to choose $N=6[2\pi^2 \left(\frac{3}{5}\right)^2\Delta_{\cal R}^2(k_0) ]^2$, where $\Delta_{\cal R}^{2}(k_0)$ is the amplitude of the primordial power spectrum of the curvature perturbations at a pivot scale $k_0$. Shapes are typically normalized such that $S(k_0,k_0,k_0)=1$.

CMB statistics are commonly described by correlations between angular moments on the sky, $a_{\ell m}$, calculated through a spherical harmonic decomposition of the photon transfer functions, $\Delta_{\ell}(k)$, integrated along the line of sight and sourced by the primordial perturbations,
\bea
a_{\ell m} &\equiv & 4\pi (-i)^{\ell} \int \frac{d^3k}{(2\pi)^3} \Delta_{\ell}(k) \Phi({\bf k}) Y_{\ell m}({\bf \hat{k}}).
\eea
The CMB 3-point correlation function is given by 
\bea \la a_{\ell_1m_1} a_{\ell_2m_2} a_{\ell_3m_3}\ra
&=& \left(\frac{2}{\pi}\right)^3\int dx x^2 \int dk_1 dk_2 dk_3 (k_1k_2k_3)^2 B_{\Phi}(k_1,k_2,k_3)  \nonumber \\
&& \times\Delta_{\ell_1}(k_1)\Delta_{\ell_2}(k_2) \Delta_{\ell_3}(k_3)  j_{\ell_1}(k_1x) j_{\ell_2}(k_2x)  j_{\ell_3}(k_3x)  
\nonumber \\
&&\times \int d\Omega_{{\bf \hat{x}}} Y_{\ell_1m_1}({\bf\hat{x}}) Y_{\ell_2m_2}({\bf\hat{x}}) Y_{\ell_3m_3}({\bf\hat{x}}).
\eea
To perform the integrals over $k$ and $x$, we use the CAMB\footnote{http://camb.info} code \cite{Lewis:1999bs}, which uses the line of sight approximation \cite{Seljak:1996is} to calculate the photon transfer functions $\Delta_{\ell}$.

We  consider purely isotropic bispectra for which the integral over $\int d\Omega_{{\bf \hat{x}}} $ is a separable geometrical factor called the Gaunt integral. The properties of the Gaunt integral require that the non-zero correlations  have $\ell_1$, $\ell_2$, and $\ell_3$ satisfying an even sum $\ell_1+\ell_2+\ell_3$ and  $|\ell_1-\ell_2|\le\ell_3\le\ell_1+\ell_2$ for $\ell_1$, $\ell_2 \le \ell_3$. Under the assumption of isotropy, the angle-averaged angular bispectrum is the 3-point analogue to the $C_{\ell}$,
\bea
 B_{\ell_1\ell_2\ell_3} &=&\sum_{m_i} \left(\begin{array}{ccc}\ell_1 & \ell_2 & \ell_3 \\ m_1 & m_2  & m_3\end{array}\right)\la a_{\ell_1m_1} a_{\ell_2m_2} a_{\ell_3m_3}\ra,
\eea
where the bracketed term is the Wigner-3j symbol.  To further separate out a purely geometrical factor from the angular-averaged bispectrum, it is convenient to work with the reduced bispectrum $b_{\ell_1\ell_2\ell_3}$, 
\bea \label{eq:cmbbisp}
B_{\ell_1\ell_2\ell_3}  &=& \sqrt{\frac{(2\ell_1+1)(2\ell_2+1)(2\ell_3+1)}{4\pi}}\left(\begin{array}{ccc}\ell_1 & \ell_2 & \ell_3 \\ 0 & 0  & 0 \end{array}\right)b_{\ell_1\ell_2\ell_3},
\eea
so that
\bea \label{eq:reducedbisp}
b_{\ell_1\ell_2\ell_3}  &=&  \left(\frac{2}{\pi}\right)^3\int dx x^2 \int dk_1 dk_2 dk_3 (k_1k_2k_3)^2 B_{\Phi}(k_1,k_2,k_3)
\nonumber \\
&& \times\Delta_{\ell_1}(k_1)\Delta_{\ell_2}(k_2) \Delta_{\ell_3}(k_3)  j_{\ell_1}(k_1x) j_{\ell_2}(k_2x)  j_{\ell_3}(k_3x).
\eea
Here $x$ is a dummy variable which should be integrated between zero and infinity. We note that in an analogous evaluation of $C_{\ell}$, $x$ has a physical interpretation as the comoving distance to the surface of last scattering. One might assume therefore that in the 3-point integral the upper limit $x_{max}$ could be set to $(\tau_0-\tau_{rec})$.  However, as others have previously also commented \cite{Senatore:2009gt}, the integral over $x$ arises out of rewriting the delta function in \eqref{eq:defBphi} as an integral over a product of Bessel functions.  Numerically, for a general bispectrum we find the value of $x_{max}$ ensuring the required degree of convergence is $\ell$-dependent, and typically needs to be greater than $(\tau_0-\tau_{rec})$.

\subsection{Shape similarity}
\label{sec:similarity}

The degree to which a bispectrum $B$ is theoretically similar  to another, $B'$, can be quantified by a $k$-space correlation coefficient, or `$k$-space cosine', corr$_k$, integrated over the $k$-space tetrapyd volume ${\cal V}$ with weight $w$  \cite{Babich:2004gb,Fergusson:2009nv},
\bea
\langle S, S'\rangle&\equiv &  \int_{\cal V} S(k_1,k_2,k_3)  S'(k_1,k_2,k_3)  w(k_1,k_2,k_3) dk_1 dk_2 dk_3, \label{eq:tetrapyd}
\\
\mathrm{corr_k}(S,S') &\equiv&  \frac{\langle S,S'\rangle_k}{\sqrt{\langle S',S'\rangle_k \langle S,S\rangle_k}}.\label{eq:kcorr}
\eea
An analogous statistic describing the similarity of two bispectra in multipole space can be quantified by their $\ell$-space correlation coefficient, or `$\ell$-space cosine', corr$_\ell$ \cite{Fergusson:2009nv}, 
\bea
\langle S,S'\rangle_\ell&\equiv& \sum_{\ell_1 \ell_2 \ell_3} \frac{B_{\ell_1 \ell_2 \ell_3} B'_{\ell_1 \ell_2 \ell_3}}{C_{\ell_1}C_{\ell_2}C_{\ell_3}}
\\
\mathrm{corr}_{\ell}(B,B') &\equiv&\frac{\langle S, S'\rangle_{\ell}}{\sqrt{\langle S,S\rangle_{\ell} \langle S',S'\rangle}_{\ell}}.\label{eq:ellcorr}
\eea

These correlation statistics are frequently used within non-Gaussian shape studies to quantify how well a template matches a given shape.
A template can be obtained, given a set of $n$ basis shapes, in our case $\{{\mathcal K}_n\}$, by first applying the Gram-Schmidt algorithm  to give a set of orthonormal basis functions $\{{\mathcal R}_n\}$ (in either $k$ or $\ell$ space),
\bea
{\cal R}_{0}' &=& {\cal K}_{0} \\
{\cal R}_{n+1}' &=& {\cal K}_{n+1} - \sum_{i=0}^n \langle {\cal K}_{n+1},{\cal R}_i \rangle \\
{\cal R}_j &=& \dfrac{{\cal R}_j'}{\sqrt{\langle {\cal R}_j',{\cal R}_j' \rangle}}
 = \sum_{i=0}^j \lambda_{ji} {\cal K}_i,
\eea
where the last line defines the matrix $\lambda$.  The new basis can then be used to create a matched template for a specific non-separable shape $S$, 
\bea
S_{template}^{(n)} &=&  \sum_{i=0}^{n}\langle {\cal R}_{i}, S\rangle {\cal R}_{i} =  \sum_{i=0}^{n}\alpha_i{\cal R}_{i} .\label{eq:template}
\eea

We note that in general the classical Gram-Schmidt algorithm can be numerically unstable, resulting in $\{{\cal R}_n\}$ that are not exactly orthogonal.  This issue can be abated by implementing the well-known modified Gram-Schmidt algorithm, and any numerical issues that may remain can be checked by verifying that all of the $\{{\cal R}_n\}$ are orthogonal to each other, i.e. $\langle {\cal R}_{i},{\cal R}_{j} \rangle = \delta_{ij}$.  For our case this was true to within a few $\times 10^{-6}$ at worst, across the first 7 modes.  In addition, a faster check can be conducted for those shapes for which the coefficients on the original basis are known, since computing $\alpha\lambda$ should return the input coefficients on the $\{{\cal K}_n\}$ basis.  We verified that this was the case for the equilateral, orthogonal, and enfolded templates, with the worst coefficients being off by a fractional error of a few $\times 10^{-5}$.  We find this accuracy is more than sufficient for the forecasting analyses to constrain shape measurements with upcoming surveys as we discuss in section 3.

The efficacy of the template can be quantified by the cumulative cosine, $\mbox{corr}(S,S_{template}^{(n)})$ as in (\ref{eq:kcorr}) or (\ref{eq:ellcorr}).
A high correlation coefficient signals a good fit. If this cosine is close to one then the two shapes are sufficiently alike that one might expect constraints on the amplitude of $B$ can be taken as constraints on $B$' as well, without having to do a separate analysis of the data.  If on the other hand, the cosine is low, then it is likely that separate analyses of the data are needed for $B$ and $B$', because a template for $B$ will not be able to pick out a non-zero signal for $B$', and vice versa.   We note though that while a template may have a large cosine with the shape, this does not automatically mean that the template will be able to accurately model a correlation between the true shape and a third shape, with which it is not similar. The extreme example of this would be if the third shape were exactly proportional to the discrepancy between the template and shape. In constructing a template, if this was a concern, one might want to tailor it to the purpose by altering the weight in the Gram-Schmidt decomposition to ensure a minimization of the covariance between the shape and template over a given region of ($k$ or $\ell$) configuration space in which a third shape was relevant. 

While it is extremely useful to establish the similarity of shapes, it is the converse of this, how well two shapes can be distinguished from one another, using data, that is the main focus of this work. This provides a motivation to consider an efficient way to generate $\ell$-space bispectra explicitly by creating templates described by basis functions separable in $k_1,k_2$ and $k_3$ as we discuss below. To do this, in sections \ref{sec:basis} and \ref{sec:motivation} we develop a framework to describe the possible degrees of freedom that a general shape might have under a specific theoretical prior.

We note that corr$_k$ and corr$_{\ell}$ represent simplified correlation statistics that purely take into account the cosmic variance limitations.
Neither statistic, as they are written above, takes into account the noise, sky coverage or resolution characteristics of a particular survey.  
As described in \cite{Fergusson:2009nv}, corr$_{\ell}$ can be modified to include these experimental effects by changing the weighted sum over $\ell_1, \ell_2, \ell_3$ to reflect the measurement covariance matrix.  The modified corr$_{\ell}$ is then a refined, survey-dependent extension of \eqref{eq:ellcorr} that tailors the correlation statistic to reflect the observational, rather than intrinsic, distinguishability of shapes.  
Distinguishability between several shapes can be done by conducting a Fisher or $\chi^2$ analysis that includes the measurement covariance matrix \cite{Komatsu:2001rj}.
In section \ref{analysis}, we perform a Fisher analysis and use the correlation statistics, including experimental effects such as instrument noise, beam size, and incomplete sky coverage, to quantify how well upcoming surveys might distinguish one shape from another.

\subsection{A new separable basis, ${\cal K}_n$}
\label{sec:basis}

The 4-dimensional integral over the product of highly oscillatory functions given in \eqref{eq:reducedbisp} is computationally intensive.  This has been a barrier to efficiently calculating observational predictions for the CMB bispectrum. As a result many studies have focused on models for which the primordial bispectrum can be written as, or well-approximated by, a separable (symmetric) function of $(k_1,k_2,k_3)$, 
\bea
S(k_1,k_2,k_3) = f(k_1)g(k_2)h(k_3) + \mbox{ cyclic perms},
\eea
such that the 3-dimensional integral over $k_1,k_2$, and $k_3$ in \eqref{eq:reducedbisp} is reduced to a product of three 1-dimensional integrals.

In \cite{Fergusson:2009nv} the authors proposed a way to reduce the computation time for general models by expanding the shape in terms of a separable basis,
\bea
S(k_1,k_2,k_3) &=& \sum_i \alpha_i {\cal Q}_i(k_1,k_2,k_3).
\eea
Each $\mathcal{Q}_i$ is constructed from symmetrized products of three 1-dimensional polynomials of $k_1$, $k_2$, and $k_3$,
\be \label{FSQn}
	\mathcal{Q}_n(k_1,k_2,k_3) \propto \sum_{i=0}^p c_{pi} \sum_{j=0}^r c_{rj} \sum_{k=0}^s c_{sk} \left( k_1^i k_2^j k_3^k + k_1^i k_2^k k_3^j + 4 \mbox{ perms} \right),
\ee
where $n$ maps onto a combination of $\{p,r,s\} \ge 0$, and $c_{pi}$, $c_{rj}$, $c_{sk}$ are constants. Using the ${\cal Q}_n$ basis, the equilateral template can be reconstructed to 98\% accuracy (according to the cumulative cosine) using 6 basis functions \cite{Fergusson:2009nv}, while other shapes motivated by single-field inflation models can require 20 or more mode functions to get $>95\%$ convergence \cite{Battefeld:2011ut}.

An analysis of data to constrain the bispectrum depends not only on the uncertainties inherent in the data itself, but also the theoretical priors determining the model being compared with the data. The choice of a separable basis set to describe the theory is therefore also influenced by this prior. An analysis allowing the primordial bispectrum to take any form (i.e. with no shape prior on the forms of the separable functions $f_i,g_i,h_i$) would use a discrete set of $k$-space bins to describe the uncorrelated amplitudes at each scale and configuration. In such a scenario, no theoretical prior is applied and the constraints on the bispectrum are simply those determined by the data. In studying theoretically motivated models of inflation, however, there can be broad or specific characteristics of the bispectra that it would be reasonable to impose in conjunction with the data that suggest a form for the separable basis functions. For example, a Fourier basis may be more efficient than a polynomial one for describing bispectra with oscillatory features \cite{Meerburg:2010ks}.  The two minimal assumptions we consider here as theoretical priors are that the bispectrum i) has a roughly monotonically changing amplitude as a function of scale, and ii) like the power spectrum, it is nearly scale-invariant. 

The polynomial basis of \cite{Fergusson:2009nv} does not naturally confine shapes to these two common theoretical properties of bispectra. Firstly, the polynomial basis does not naturally restrict itself to scale-invariant shapes, because $i+j+k \geq 0$ in \eqref{FSQn}; resulting sums of the basis functions are thus scale-dependent in general.  Most theoretically-motivated bispectra in the literature, however, are nearly scale-invariant, with $i+j+k\approx0$  (see \cite{Chen:2010xka} for a review). Such shapes can be reduced to functions of two variables, $k_1/k_3$ and $k_2/k_3$.  There are exceptions to this, of course, such as non-Gaussianity from particle production \cite{Barnaby:2010ke} or from features in the inflationary potential \cite{Chen:2006xjb,Chen:2008wn}. These models can strongly deviate from scale-invariance because  modes leaving the horizon at a specific moment, when  particle production is occurring or a feature in the potential is important, are preferentially populated. Secondly, different types of theoretical mechanisms generating bispectra predict different divergence properties  in the squeezed limit, where $k_1 \ll k_2 \approx k_3$. We consider the squeezed limit as $k_1=k_{\ell}$, the long wavelength mode, and $k_2=k_3=k_s$, the short wavelength modes, so that for  scale-invariant shapes the squeezed limits purely dependent on $x_{sq}\equiv k_{\ell}/k_s$.  Single-field inflation models, through a consistency relation \cite{Maldacena:2002vr} predict the bispectrum will vanish in this limit. Local bispectra, typically arising in multi-field models, have a $x_{sq}^{-1}$ divergence, while excited states can have $x_{sq}^{-2}$ divergence. Since the powers $i$, $j$, and $k$ in \eqref{FSQn} are $\ge 0$, the $\{\mathcal{Q}_n\}$ all tend towards a constant value in the squeezed limit, and thus cannot effectively describe shapes diverging in the squeezed limit.  As a result, this basis is not immediately suited to reconstructing templates that display specific divergence behaviors in this limit, without a further prior being imposed.  For example, the compelling and well-studied local template cannot easily be recovered using the $\{\mathcal{Q}_n\}$ basis without either using many more basis functions or ignoring the divergent part of the shape that makes the local template distinct from others \cite{Battefeld:2011ut}.

In this paper we introduce a set of separable basis functions, $\{{\cal K}_{n}\}$, that efficiently describe nearly scale-invariant and potentially divergent shapes, and explicitly consider the forms of the shapes generated using this basis under various divergence constraints.  Explicitly, we consider
\bea
f_{NL} S(k_1,k_2,k_3)=  \sum_n f^n_{NL} {\cal K}_{n}(k_1,k_2,k_3).
\label{fnln}
\eea
Here $f_{NL}^n$ are expansion coefficients and $n$ again denotes a combination of powers $\{p,r,s\}$ of the wavenumbers $(k_1,k_2,k_3)$.  ${\cal K}_{n}$ is defined as
\bea \label{first Kn defined}
{\cal K}_{n}(k_1,k_2,k_3) &\equiv& \frac{1}{{\cal N}_{n} k_0^{2(n_s-1)}}\left[k_1^{p'} k_2^{r'} k_3^{s'}+ \{prs\} \operatorname{ perms}\right],
\eea
where ${\cal N}_{n}$ is the number of distinct permutations of $\{p,r,s\}$.  $p'$ is defined as
\bea
p' &=& 2+\frac{(p-2)(4-n_s) }{ 3}, \label{eq:pprime}
\eea
and similarly for $r'$ and $s'$. 

Equations \eqref{first Kn defined}--\eqref{eq:pprime} ensure that each $\mathcal{K}_{n}$ is normalized in the conventional way, with ${\cal K}_{n}(k_1,k_2,k_3)=1$ at $k_1=k_2=k_3=k_0$. In the scale-invariant case where $n_s=1$, ${\cal K}_{n}$ only depends on $k_1/k_3$ and $k_2/k_3$, and ${\cal K}_{n}(k_1,k_2,k_3)=1$ for all $k_1=k_2=k_3$. 

To allow for potentially divergent shapes, we allow the powers $\{p,r,s\}$ to be negative as well as positive, and to make each $\mathcal{K}_{n}$ nearly scale-invariant, we require the powers to satisfy $p+r+s=0$.  

Each shape has a well-determined behavior in the squeezed limit,
\bea
{\cal K}_{n}^{sq}= \sum_{m=R}^{1} d_{nm} x_{sq}^{m} + \mathcal{O}(x_{sq}^2).
\eea
The set of $\{\mathcal{K}_{n}\}$ with the allowed combinations of $\{p,r,s\}$ is  given in Table \ref{tab:Kn} along with their divergence properties. The basis modes are also written in the equivalent short-hand notation used by \cite{Fergusson:2008ra}. Since we will use this notation to describe non-separable shapes in the next section we summarize it here:
\bea
K_{p}&=& \sum_{i=1}^{3} k_i^p,
\hspace{0.5cm} K_{pq}=\frac{1}{\Delta_{pq}}\sum_{i=1}^{3} k_i^p\sum_{j\ne i}k_j^q,
\hspace{0.5cm} K_{pqr} =\frac{1}{\Delta_{pqr}}\sum_{i=1}^{3} k_i^p\sum_{j\ne i}k_j^q\sum_{\ell\ne i,j}k_\ell^r, \label{eq:Kpqr}
\eea
with
\bea
\Delta_{pq} &=& 1+\delta_{pq},
\hspace{0.5cm} \Delta_{pqr} = \Delta_{pq}(\Delta_{qr}+\delta_{pr}).
\eea

\begin{table}[t]
\centering
\begin{tabular}{|c|c|c|c|c|c|c|c|}
 \hline
$R=$ & \multirow{2}{*}{${\cal K}_n$}& \multirow{2}{*}{$\{p,r,s\}$}   &  \multicolumn{4}{|c|}{$d_{nm}, m=$} & \multirow{2}{*}{$K_{pqr}$} 
\\ \cline{4-7}
$min(p,r,s)$& & & -3 & -2 & -1 & 0 &
\\ \hline
 0&${\cal K}_0$ & (0,0,0) &  & & &1  & $K_{000}$ 
\\  \hline
\multirow{2}{*}{-1}&${\cal K}_1$ & (-1,0,1) & & &$\frac{1}{3}$ &$\frac{1}{3}$  &$\frac{K_{12}}{K_{111}}$ 
\\
& ${\cal K}_2$ & (-1,-1,2) & & &$\frac{2}{3}$  &&$\frac{K_{3}}{K_{111}}$
\\ \hline
\multirow{4}{*}{-2}&${\cal K}_3$ & (-2,1,1) & & $\frac{1}{3}$ & && $\frac{K_{33}}{K_{222}}$ 
 \\
& ${\cal K}_4$ & (-2,0,2) & &$\frac{1}{3}$  & &$\frac{1}{3}$ &$\frac{K_{24}}{K_{222}}$ 
\\ 
&${\cal K}_5$ & (-2,-1,3) & & $\frac{1}{3}$ & $\frac{1}{3}$ & & $\frac{K_{15}}{K_{222}}$ 
\\
&${\cal K}_6$ & (-2,-2,4) & & $\frac{2}{3}$ & &&  $\frac{K_{6}}{K_{222}}$
 \\ \hline
\multirow{5}{*}{-3}&${\cal K}_7$ & (-3,1,2) &$\frac{1}{3}$ & & & &  $\frac{K_{45}}{K_{333}}$
\\
&${\cal K}_8$ & (-3,0,3) &$\frac{1}{3}$ & & &$\frac{1}{3}$  & $\frac{K_{36}}{K_{333}}$
\\
&${\cal K}_9$ & (-3,-1,4) &$\frac{1}{3}$ & & $\frac{1}{3}$ & &$\frac{K_{27}}{K_{333}}$
\\
&${\cal K}_{10}$ & (-3,-2,5) &$\frac{1}{3}$ &$\frac{1}{3}$  & & & $\frac{K_{18}}{K_{333}}$
\\
&${\cal K}_{11}$ & (-3,-3,6) &$\frac{2}{3}$ & & & &  $\frac{K_{9}}{K_{333}}$
  \\ \hline 
& etc. & && & & &
\\\hline
\end{tabular}
\caption {\label{tab:Kn}The set of $\mathcal{K}_{n}$ with the allowed combinations of $\{p,r,s\}$ that satisfy $p+r+s=0$.  The ordering of the modes is according to their divergence behavior in the squeezed limit.  The coefficients of their divergent terms in this limit, $d_{nm}$, are trivially related to the values of $\{p,r,s\}$, but are given here for convenience.  We also give the equivalent short-hand notation used in \cite{Fergusson:2008ra} and summarized in (\ref{eq:Kpqr}).}
\end{table}

\subsection{Application of the basis to shapes arising in inflationary theory}
\label{sec:motivation}

In this subsection, we illustrate the efficiency and accuracy allowed by our basis in describing shapes in the literature.  First we discuss cases involving shapes and templates which are exactly expressed in terms of, or well-approximated by, linear combinations of the first 3 modes of the basis.  Then we extend the basis to include more divergent modes, and present the basis of shapes we will use under different  divergence priors.

\subsubsection{Shapes exactly expressed in terms of $\{{\cal K}_0-{\cal K}_2\}$}

Some commonly considered templates can be exactly expressed in terms of the first three modes of the basis $\{{\cal K}_{0}-{\cal K}_{2}\}$.

The local shape, $S_{local}={\cal K}_{2}$, can be derived from a simple ansatz for describing the nonlinear contribution to the primordial curvature perturbation in real space as a local effect \cite{Komatsu:2001rj},
\be
\Phi(\mathbf{x}) = \Phi_L(\mathbf{x}) + f_{NL}\left(\Phi_L^2(\mathbf{x}) - \left<\Phi_L^2(\mathbf{x})\right> \right).
\ee
 Local shapes arise out of single-field slow-roll models, though the amplitude of the bispectrum in this case is 
predicted to be undetectably small \cite{Maldacena:2002vr,Acquaviva:2002ud}.
Large, local non-Gaussianity is predicted by a wide variety of other physically-motivated models, such as multifield inflation (e.g. curvaton scenario) \cite{Wands:2007bd,Byrnes:2010em}, (p)reheating mechanisms \cite{Enqvist:2004ey}, and ekpyrotic inflation \cite{Byrnes:2010em,Wands:2010af}.
 
The constant shape, $S_{const}=1={\cal K}_{0}$, was originally studied for its very simple form \cite{Fergusson:2008ra}. More recently the shape has been studied in the context of shapes arising from quasi-single field inflation (QSFI) models \cite{Chen:2009we,Chen:2009zp,Sefusatti:2012ye,Norena:2012yi}.  The more general shape of QSFI is discussed in more detail below.

Models with higher-derivative kinetic terms and/or non-trivial speeds of sound in the inflationary Lagrangian generally produce non-separable shapes, sensitive to the sum $k_t=k_1+k_2+k_3$ in the denominator, and thus cannot be exactly written in terms of a separable basis.   The equilateral template \cite{Creminelli:2005hu}, $S_{equil}=-2{\cal K}_0+6{\cal K}_1-3{\cal K}_2$, is widely used as a template to detect evidence of such shapes.   Examples include generalized single-field models \cite{Creminelli:2003iq,Seery:2005wm,Chen:2006nt,Cheung:2007st}, $k$-inflation \cite{ArmendarizPicon:1999rj,Garriga:1999vw,Li:2008qc}, ghost inflation \cite{ArkaniHamed:2003uz}, DBI inflation \cite{Silverstein:2003hf,Alishahiha:2004eh}, single-field non-slow roll and bimetric theories \cite{Ribeiro:2012ar,Magueijo:2010zc,Noller:2011hd}.

A general, effective field theory of inflation is dominated by contributions from two shapes \cite{Chen:2006nt},
\bea
S_{DBI}&=& \frac{3}{7 K_{111}}\left(\frac{8K_{22}}{k_t}-\frac{4K_{23}}{k_t^2}-K_3\right) , 
\\
S_{single} &=& \frac{27 K_{111}}{k_t^3}.
\eea
While each can typically be well-described by the equilateral template, a linear combination of these picking out the differences between them can yield a very different shape. This realization led to the generation of the `orthogonal' template, $S_{orth}=-8{\cal K}_0+18{\cal K}_1-9{\cal K}_2$ \cite{Senatore:2009gt}.

While inflation derived from a Bunch-Davies vacuum can be written in terms of a plane wave basis with positive $k$ modes, excited states that are not in the Bunch Davies-vacuum can have initial states with both positive and negative $k$. Models motivated by non-trivial vacuum states can produce shapes with denominators containing $k_1+k_2-k_3$ (and its permutations), rather than $k_t$ \cite{Chen:2006nt,Holman:2007na,Meerburg:2009ys,Agarwal:2012mq}. Unlike the equilateral and local templates, these shapes peak in the flattened configuration, when $k_3=k_1+k_2$. While this shape again cannot be reconstructed perfectly using separable basis functions, an ansatz proposed as a proxy to this shape can be given by $S_{enf} =  -3{\cal K}_{0}+6{\cal K}_{1}-3{\cal K}_{2}$ \cite{Meerburg:2009ys}. The shape has zero amplitude at $k_1=k_2=k_3$, making the conventional normalization at this configuration unsuitable for this template. Though flattened shapes such as this one are usually associated with generalized initial states, it is in some cases possible to obtain flattened shapes through single-field inflation \cite{Bartolo:2010bj}.

\subsubsection{Shapes well-approximated by $\{{\cal K}_0-{\cal K}_2\}$}

Non-Gaussian templates to describe single-field theories are not limited to equilateral and orthogonal shapes. Fast-roll single-field non-Gaussian models \cite{Khoury:2008wj,Noller:2011hd} retain the scale-invariant spectra but relax the condition for slow-roll inflation. \cite{Battefeld:2011ut}  showed these can be written in terms of seven constituents,  four of which are $S_{local}$, $S_{const}$, ${\mathcal K}_1$, and $S_{single}$. The remaining three constituent shapes are \footnote{Our shape, $S_n$, is related to Battefeld and Grieb's ${\mathcal A}_n$, by $S_n={\mathcal A_n}/K_{111}$.}
\bea
S_3 &=& \frac{K_{22}}{K_{111}k_t},\hspace{0.5cm}  S _4=\frac{K_{23}}{K_{111}k_t^2} , \hspace{0.5cm} S_5 =\frac{K_{6}}{K_{111}k_t^3},
\eea
all of which have significant cosines with the local template.

Other shapes exist in the literature that, while not separable, to some degree interpolate between the templates discussed above and hence can be reasonably-well described by linear combinations of $\{\mathcal{K}_{0}-{\cal K}_{2}\}$. For example, non-Bunch-Davies vacua  generate shapes that can be equilateral, local, or enfolded  \cite{Agarwal:2012mq}.

Quasi-single field (QFSI) models \cite{Chen:2009we,Chen:2009zp,Sefusatti:2012ye,Norena:2012yi} motivated by string theory and supergravity inspired inflation contain multiple fields, but the extra fields have masses comparable to the Hubble scale. These models can be well described by a family of bispectrum templates dependent on a single parameter, $\nu$, 
\bea \label{eq:qsfi_shape}
S_{QSFI}(\nu)&=& \left(\frac{3k_1k_2k_3}{k_t}\right)^{3/2}\frac{N_{\nu}\left[8k_1k_2k_3 k_t^{-3}\right]}{N_{\nu}[8/27]},
\eea
where $N_{\nu}$ is the Neumann function of order $\nu$. This shape interpolates between the constant and local templates. Another set of models that combine multiple fields and higher-derivative terms \cite{Arroja:2008yy,Langlois:2008qf,RenauxPetel:2009sj} also generate configurations that interpolate between standard shapes, spanning the local and equilateral templates.  

\begin{figure}[t]
    \centering
        \includegraphics[width=1.02\textwidth]{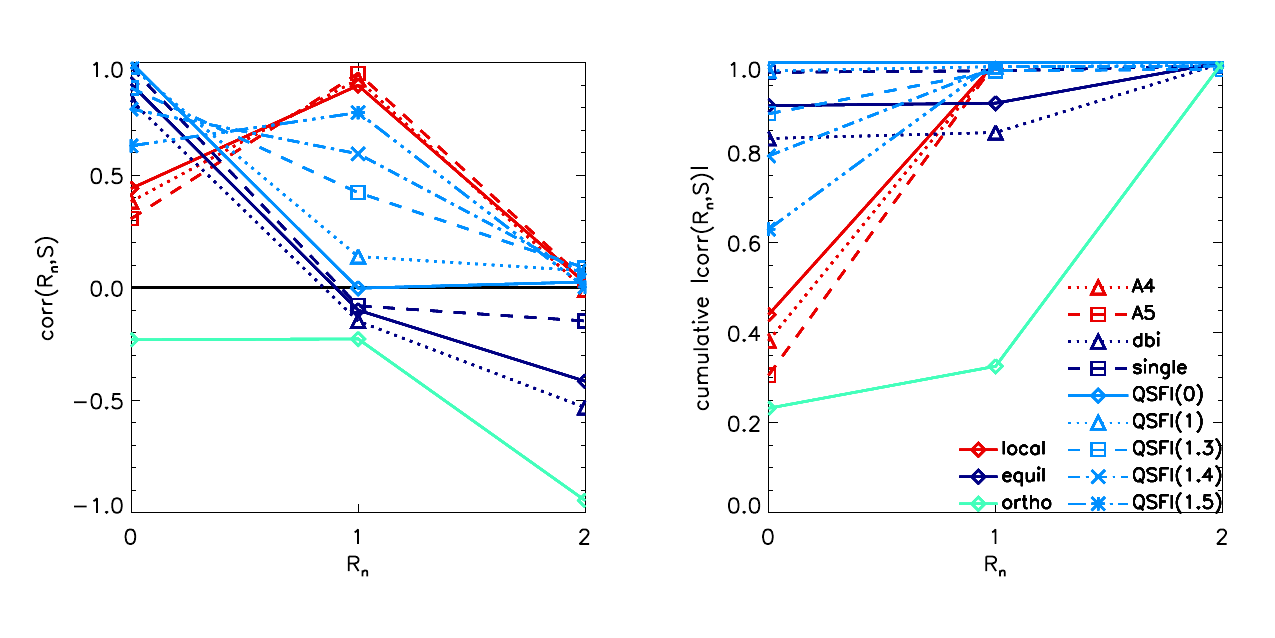}
        \caption{The application of the  separable basis to describe shapes motivated by  theoretically distinct  models that span a wide array of configurations in $k$-space.   The local, equilateral, and orthogonal templates are explicitly separable, and are included only for reference. The other shapes are not separable but are approximated by templates using linear combinations of 
        $\{\mathcal{K}_0,\mathcal{K}_1,\mathcal{K}_2\}$. We construct an orthonormal basis, $\mathcal{R}_n$, in $k$-space with uniform weighting, using a Gram-Schmidt decomposition for $\mathcal{K}_n$, for $0\le n\le 2$, starting with $n=0$ .  [Left panel] The cosines between each shape and the $\mathcal{R}_n$. [Right panel] The cumulative cosine between a constructed template  $\sum_{i=0}^n c_i\mathcal{R}_i$ and the true shape.  }\label{fig:shape-corr}
       \end{figure}

We use  the  basis modes, $\{\mathcal{K}_{0}-{\cal K}_{2}\}$ to create templates for these non-separable shapes, $S_{3-5}$, $S_{DBI}$, $S_{single}$, and $S_{QFSI}(\nu)$. To demonstrate this, we  generate an orthonormal basis $\{{\mathcal  R}_{n}\}$ using the Gram-Schmidt algorithm in $k$-space, taking ${\mathcal R}_{0}={\mathcal K}_{0}$, and create a template $S_{template}=\sum_{i=0}^{n}\alpha_n {\cal R}_{n}$ as in (\ref{eq:template}) that reduces the covariance between the shape and template. The effectiveness of the template's fit can be quantified by the cumulative cosine. In Figure \ref{fig:shape-corr} we show how the shapes discussed above can be well modeled by templates using linear combinations of the 
$\{\mathcal{K}_0-\mathcal{K}_2\}$
templates.  In each case the cumulative cosine for the template and shape exceeds 0.98.  
   
\subsubsection{Shapes well-approximated by more divergent $\{{\cal K}_0-{\cal K}_n\}$}

There are two strong motivations to extend template design beyond these three core templates. 
Firstly, expansions using the first three templates do not necessarily ensure that theoretical priors on the divergence properties are satisfied by the template.  An example of this is the  consistency relation that requires shapes of single-field inflation to vanish in the squeezed limit \cite{Maldacena:2002vr}. However, the orthogonal and enfolded templates constructed to describe single-field shapes tend toward a constant value in the squeezed limit. \cite{Senatore:2009gt} proposed an orthogonal template, $S_{ortho(2)}$,  and \cite{Creminelli:2010qf} an  enfolded template, $S_{enf(2)}$, that are somewhat more complex, using linear combinations of shapes that  diverge as ${x_{sq}}^{-2}$, but they have the benefit of showing the correct divergence properties and more accurately reproducing the original non-separable shape. They can be written in terms of the ${\cal K}_{n}$ modes as
\bea
S_{ortho(2)} &=&(1+p)S_{equil} -p\left( \frac{2}{9} {\cal K}_{0}   + \frac{8}{3}{\cal K}_{1} - 2 {\cal K}_{2} + \frac{20}{9} {\cal K}_{3}  - \frac{10}{3} {\cal K}_{4}   + \frac{4}{3} {\cal K}_{5} -  \frac{1}{9} {\cal K}_{6}\right)
\\
S_{enf(2)} &=&(1+p)S_{equil}-p\left(\frac{ 6}{5}{\cal K}_{0}+\frac{16}{5}{\cal K}_{3}-\frac{18}{5}{\cal K}_{4}+\frac{1}{5}{\cal K}_{6}\right),
\eea
where $p$ is a variable chosen to maximize the template's fit to the physical shape. 

Using our basis we can generalize this approach and write down classes of templates, denoted $S_{[R,r]}$, constructed from  basis modes with maximal divergence $R$ that in the squeezed limit diverge as $x_{sq}^{r}$, where $r \geq R$. In general, a shape written in terms of the basis will have a squeezed limit behavior given by
\bea
S_{sq} = \alpha_n\sum_{m=R}^1 d_{nm}x_{sq}^{m} + \mathcal{O}(x_{sq}^2),
\eea
with $d_{nm}$ summarized in Table \ref{tab:Kn}. We find $S_{[R,r]}$  can be written in terms of an irreducible set of shapes given in Table \ref{tab:subshapes} for which $\alpha_n d_{nm}=0$ for $R\le m<r$. 

\begin{table}
\centering
\begin{tabular}{|c||l|l|l|l|}
\hline
\multirow{2}{*}{$R$} & \multicolumn{4}{|c|}{$r$}
\\
\cline{2-5}
& -2 & -1 & 0 & 1 
\\ \hline
\multirow{2}{*}{-1} &  & ${\cal K}_{2}=S_{local}$ & ${\cal K}_{0}$  & $-2 {\cal K}_{0}  + 6{\cal K}_{1}+3{\cal K}_{2}=S_{equil}$
\\
&  &  & $2 {\cal K}_{1}-{\cal K}_{2}$ &
\\ \hline
\multirow{3}{*}{-2} & ${\cal K}_6$ &  $2{\cal K}_{5}-{\cal K}_{6} $& $2{\cal K}_{4}-{\cal K}_{6} $ & $ {\cal K}_{0}+3{\cal K}_{3}-3{\cal K}_{4}$
\\
 &  & && ${\cal K}_{2}+2{\cal K}_{3}-2{\cal K}_{5}$
 \\
 &   & & & $2{\cal K}_{3}-{\cal K}_{6}$
 \\ \hline
\end{tabular}
\caption{Shapes constructed from basis modes with maximal divergence $x_{sq}^{R}$ ($R<0$) which, through cancellations of the divergent terms, have a squeezed limit that diverges as $x_{sq}^{r}$. These represent an irreducible set of component shapes, for each value of $R$, from which general, scale invariant, separable shapes can be constructed.}\label{tab:subshapes}
\end{table}

$S_{local}$ and $S_{equil}$ are the only shapes constructed from $R=-1$ modes that respectively have $-1$ and vanishing divergence. There are an infinite set of shapes, however, with constant divergence described by
$\beta{\cal K}_{0} + (1-\beta)(2{\cal K}_{1}-{\cal K}_{2})$ where $\beta$ is free parameter which could take any value except $\beta=-2$, for which the equilateral template is recovered. 
Instead of varying the parameter $\beta$, we could instead select a value of $\beta$ to generate a template from the set.  $\beta=-8$  corresponds to the orthogonal template chosen by \cite{Senatore:2009gt} to maximize the resulting shape's orthogonality with $S_{local}$ and $S_{equil}$. We could then choose to write general shapes in terms of linear combinations of $\{S_{equil}, S_{ortho},S_{local}\}$, rather than $\alpha_n{\cal K}_n$,
\bea
S_{[-1,0]}&=&\alpha_E S_{equil}+\alpha_O S_{ortho},\label{eq:Sm10}
\\
S_{[-1,-1]}&=&\alpha_E S_{equil}+\alpha_O S_{ortho} + \alpha_LS_{local}.
\eea
If these are the only shapes being used,  the normalization constraint $S_{[R,-r]}(k_0,k_0,k_0)=1$ fixes one $\alpha$ coefficient.

We can extend this approach to include basis modes that diverge as ${x_{sq}}^{-2}$, 
\bea
S_{[-2,1]} &=& \alpha_E S_{equil} +\alpha_O(S_{ortho}+6{\cal K}_{4}-6{\cal K}_{3}) +\alpha_L(S_{local}+2{\cal K}_{3}-2{\cal K}_{5})\nonumber \label{eq:Sm2p1}
\\
&&+(1-\alpha_E-\alpha_O-\alpha_L)(2{\cal K}_{3}-{\cal K}_{6}),
\\
S_{[-2,0]}&=&\alpha_E S_{equil}+\alpha_O S_{ortho}+\alpha_L(S_{local}+2{\cal K}_{3}-2{\cal K}_{5})  +\beta_3(2{\cal K}_{3}-{\cal K}_{6})\nonumber \\ &&+(1-\beta_3-\alpha_L-\alpha_E-\alpha_O)(2{\cal K}_{4}-{\cal K}_{6})
\\
S_{[-2,-1]}&=&\alpha_E S_{equil}+\alpha_O S_{ortho}+\alpha_L S_{local} +\beta_3(2{\cal K}_{3}-{\cal K}_{6})+\beta_4(2{\cal K}_{4}-{\cal K}_{6})\nonumber \\ &&+(1-\beta_3-\beta_4-\alpha_L-\alpha_E-\alpha_O)(2{\cal K}_{5}-{\cal K}_{6}),
\\
S_{[-2,-2]} &=& \alpha_E S_{equil}+\alpha_O S_{ortho}+\alpha_L S_{local} +\beta_3(2{\cal K}_{3}-{\cal K}_{6})+\beta_4(2{\cal K}_{4}-{\cal K}_{6})\nonumber \\ &&+\beta_5(2{\cal K}_{5}-{\cal K}_{6})+(1-\beta_3-\beta_4-\beta_5-\alpha_L-\alpha_E-\alpha_O){\cal K}_{6}. \label{eq:Sm2m2}
\eea
To tie this general approach to specific shapes in the literature, $S_{ortho(2)}$ and $S_{enf(2)}$ can be written in this form by the following choice of coefficients:
\be \label{eq:S_ortho(2)}
S_{ortho(2)} = (1+p)S_{equil} -pS_{[-2,1]}\left[\alpha_E=-\frac{19}{9},\alpha_O=\frac{5}{9},\alpha_L=\frac{2}{3} \right]
\ee
\be \label{eq:S_enf(2)}
S_{enf(2)} = (1+p)S_{equil}- p S_{[-2,1]}\left[\alpha_E=\frac{9}{5},\alpha_O=-\frac{3}{5},\alpha_L=0 \right].
\ee

The inclusion of extra basis shapes can be particularly important when the shape has undulations and is not just a smooth monotonic function.  Shapes arising out of Galileon inflation are a good example of this. Imposing a Galilean symmetry on a single-field inflation model \cite{Nicolis:2008in,Burrage:2010cu,Creminelli:2010qf,Ribeiro:2011ax} gives rise to a non-Gaussian shape generated by three cubic interaction terms in the inflaton Lagrangian. While the shapes associated with each of these three operators, individually, are well-approximated by $S_{equil}$ and $S_{enf(2)}$, there exist combinations of them for which the resulting Galileon shape has little overlap with any of the shapes we have mentioned so far. Non-separable templates for Galileon inflation have been developed in \cite{Creminelli:2010qf} and \cite{Ribeiro:2011ax} which have high cosines both with the underlying shape and each other.

For illustrative purposes, we consider the shape presented in \cite{Creminelli:2010qf}, based on equations (26)-(28) of this reference. When we use the Gram-Schmidt decomposition to construct a template with only the first three modes, we find a poor fit with a cumulative cosine of only 0.13. The Galileon shape derives from a single-field action and a Bunch-Davies vacuum so theoretical consistency requires that it vanishes in the squeezed limit.  Motivated by this, if we fit the Galileon model using the 4 shapes in $S_{[-2,1]}$, we obtain a template with a cosine of 0.93. This reconstruction is not improved if we allow an unconstrained combination of the seven ${\cal K}_{0}-{\cal K}_{6}$ modes.

We can extend our approach to $R=-3$ modes, and for example consider the following general shape that vanishes in the squeezed limit:
\bea
S_{[-3,1]}&=& \alpha_E S_{equil} +\alpha_O(S_{ortho}+6{\cal K}_{4}-6{\cal K}_{3}) +\alpha_L(S_{local}-2{\cal K}_{5}+2{\cal K}_{3})+ \beta_3(2{\cal K}_3-{\cal K}_6)\nonumber
\\
&& + \beta_7(2{\cal K}_7-{\cal K}_{11})+ \beta_8(S_{ortho}+6{\cal K}_{8}-6{\cal K}_{7})+ \beta_9(S_{local}-2{\cal K}_{9}+2{\cal K}_{7})\nonumber
\\
&&+(1-\alpha_E-\alpha_O-\alpha_L-\beta_3-\beta_4-\beta_7-\beta_8-\beta_9)({\cal K}_{6}-2{\cal K}_{10}+2{\cal K}_{7}).
\eea
Fitting these eight distinct shapes in $S_{[-3,1]}$ to the Galileon shape, we obtain an improved template with cosine of 0.99.

The second reason to consider a basis including more divergent terms is that some inflationary scenarios, such as excited initial states and warm inflation, in which inflation occurs in a warm radiation bath \cite{Gupta:2002kn,Moss:2007cv,LopezNacir:2011kk} (see \cite{Berera:2008ar} for a review), can give rise to shapes that are more divergent than the local shape, with an overall divergence of ${x_{sq}}^{-2}$. This would suggest using an unconstrained combination of ${\cal K}_{0}-{\cal K}_{6}$ modes, or using constrained combinations of the $R=-3$ modes for which the $x_{sq}^{-3}$ divergent term vanishes. One such example of this is a template for warm inflation proposed by \cite{Battefeld:2011ut},
\bea
S_{warm} &=& \mathcal{K}_{2}+\mathcal{K}_{7}-\mathcal{K}_{9}.
\eea

The realization that the  differences between similar shapes can be important and provide an additional insight into the underlying model, implies that we should not just compare a small number of templates to the data. It is reasonable to extend beyond this and create more refined templates, sensitive to more than just properties that models have in common with the equilateral, orthogonal, and local templates. 
 
\subsection{Shape classification and depiction}
\label{sec:classification}

The models discussed in the previous section reflect only a sample of the wide range of non-Gaussian inflationary shapes arising in the literature. Putting a coarse filter on their properties, one might characterize them using three descriptors: i) their divergence in the squeezed limit,  ii) how many modes it takes to accurately describe them,  and iii) the ``family'' to which they belong. 
 
Many of the physical shapes tend to be grouped in terms of a ``family'' resemblance to an existing template, reflecting the type of configurations of triangles with side lengths $k_1,k_2$, and $k_3$ where the shapes have most of their power \cite{Fergusson:2008ra,Liguori:2010hx}. For scale invariant shapes this is equivalent to studying the distribution of power over the space $\{\frac{k_1}{k_3},\frac{k_2}{k_3}\}$ for a fixed $k_3>k_1,k_2$. This space can be pictorially represented by a triangle with sides $0\le\frac{k_1}{k_3}\le 1$ and $1/2\le \frac{k_2}{k_3} \le 1$. We introduce it here in the context of the shapes already discussed, because we use this format to present some of our forecasting results. 

\begin{figure}[t]
    \centering
    \subfloat[subfig1 text][$S_{local}$]
        {
        \includegraphics[width=0.33\textwidth]{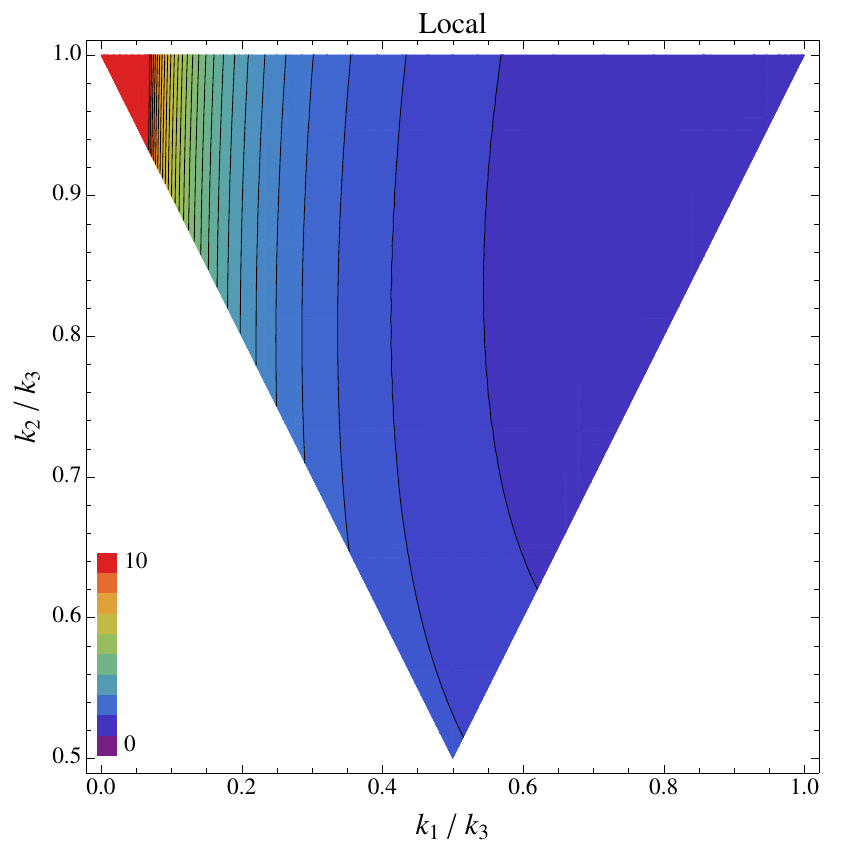}
        }
    \subfloat[subfig2 text][$S_{equil}$]
        {
        \includegraphics[width=0.33\textwidth]{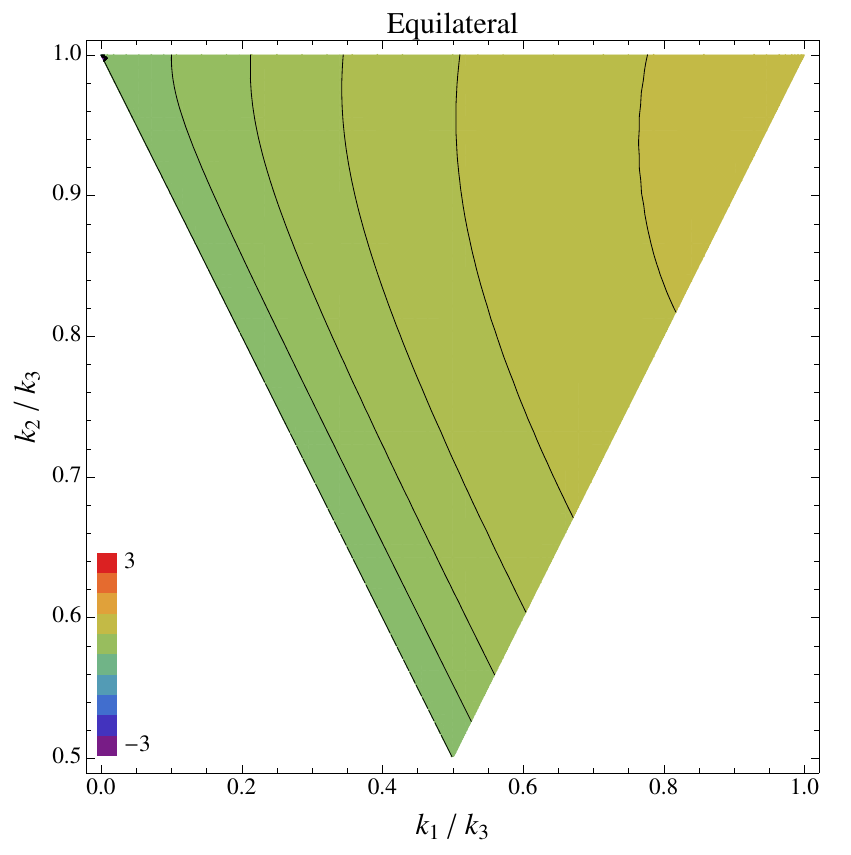}
        }
                           \subfloat[subfig4 text][$S_{enf}$]
        {
        \includegraphics[width=0.33\textwidth]{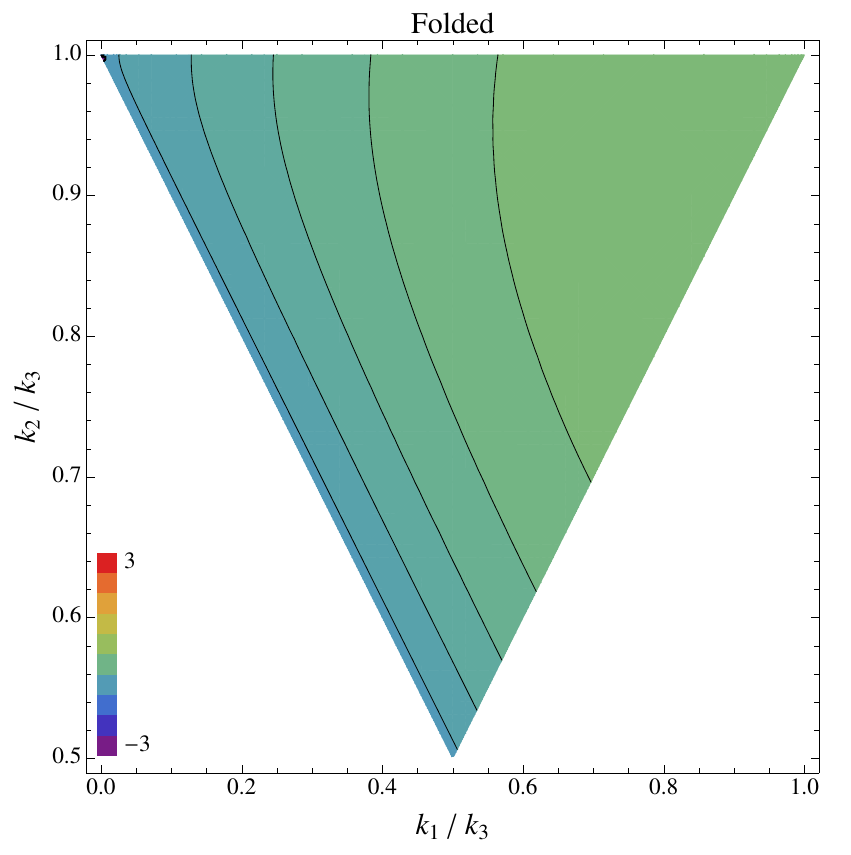}
        }
                        \\
                            \subfloat[subfig3 text][$S_{ortho(2)}$]
        {
        \includegraphics[width=0.33\textwidth]{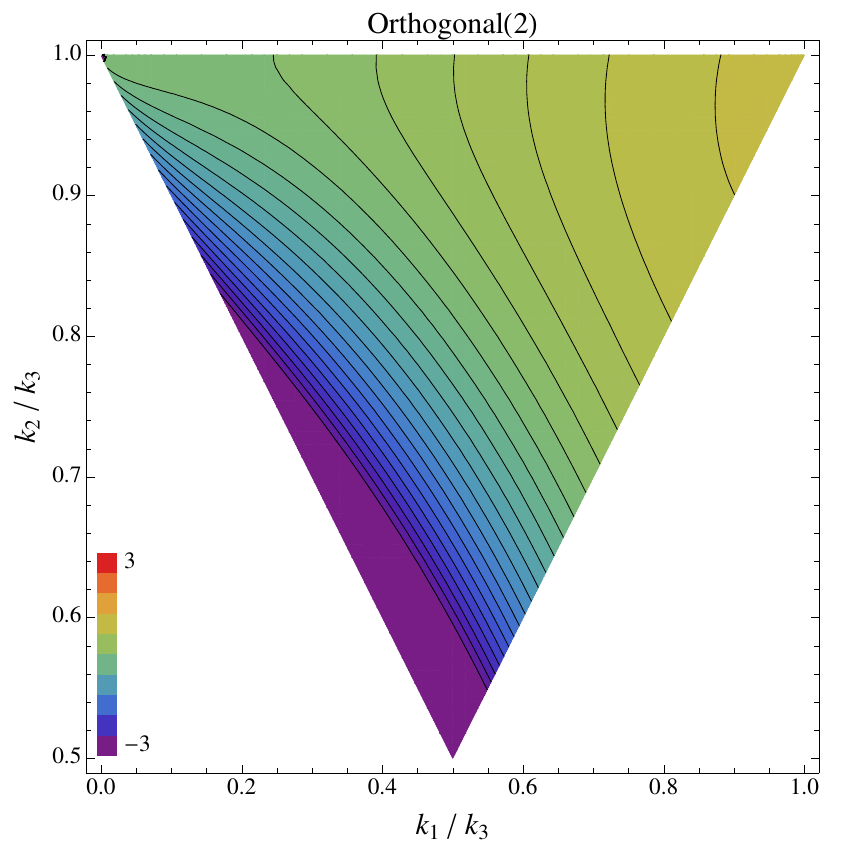}
        }
       \subfloat[subfig5 text][$S_{QSFI}(\nu=1.3)$ ]
        {
        \includegraphics[width=0.33\textwidth]{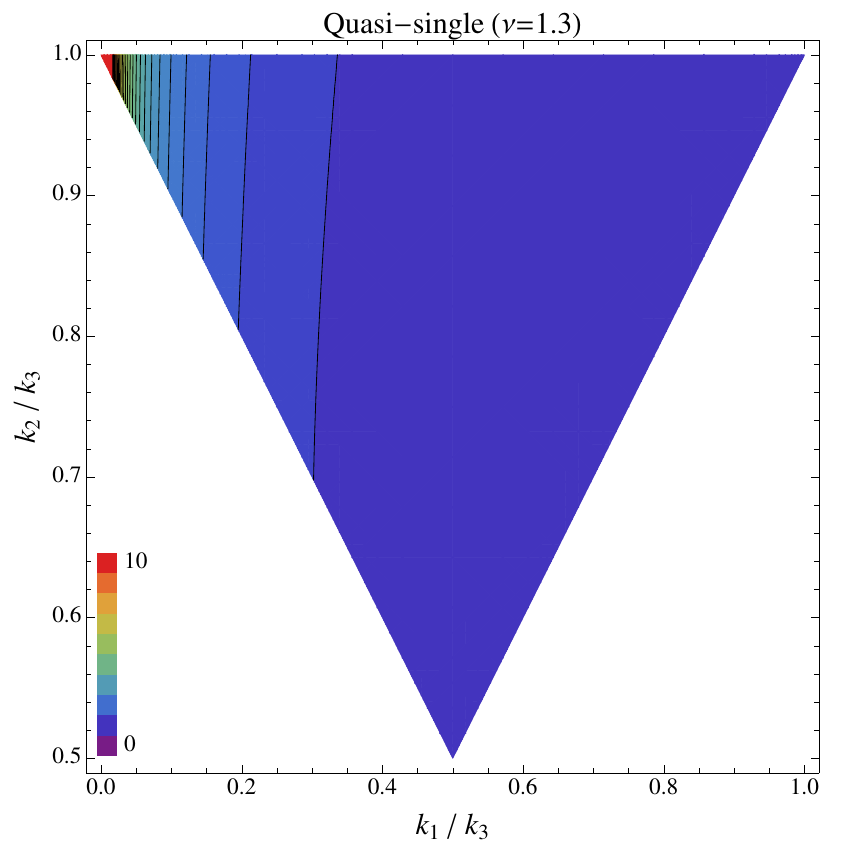}
        }
      \subfloat[subfig6 text][$S_{Galileon}$ ]
        {
        \includegraphics[width=0.33\textwidth]{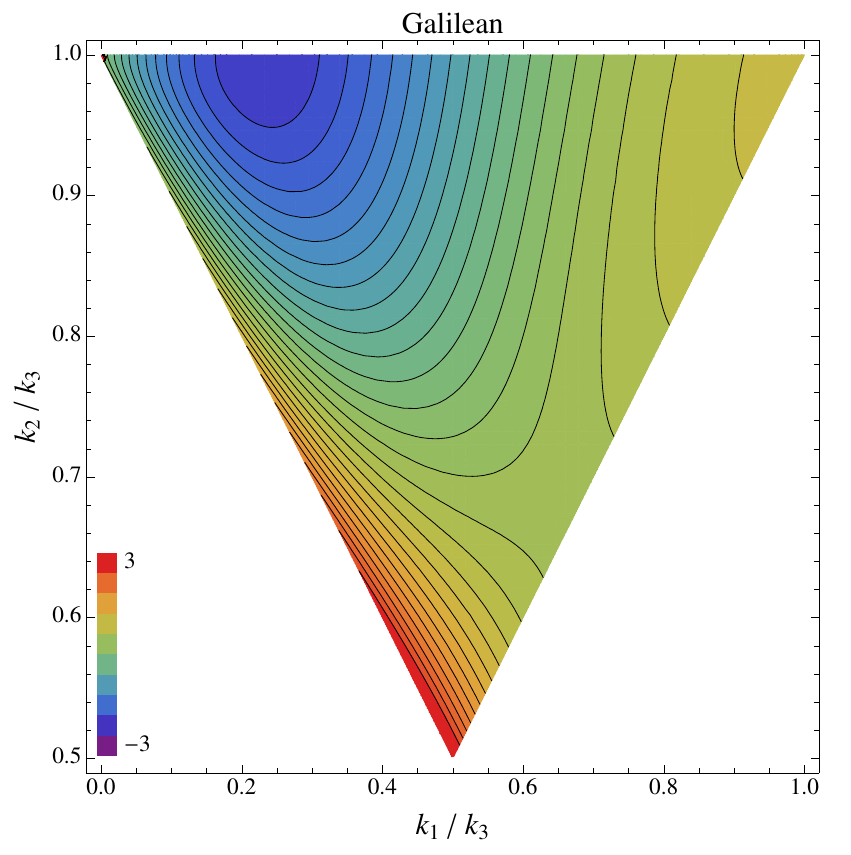}
}
\\
                            \subfloat[subfig3 text][$S_{ortho}+6{\cal K}_{4}-6{\cal K}_{3}$]
        {
        \includegraphics[width=0.33\textwidth]{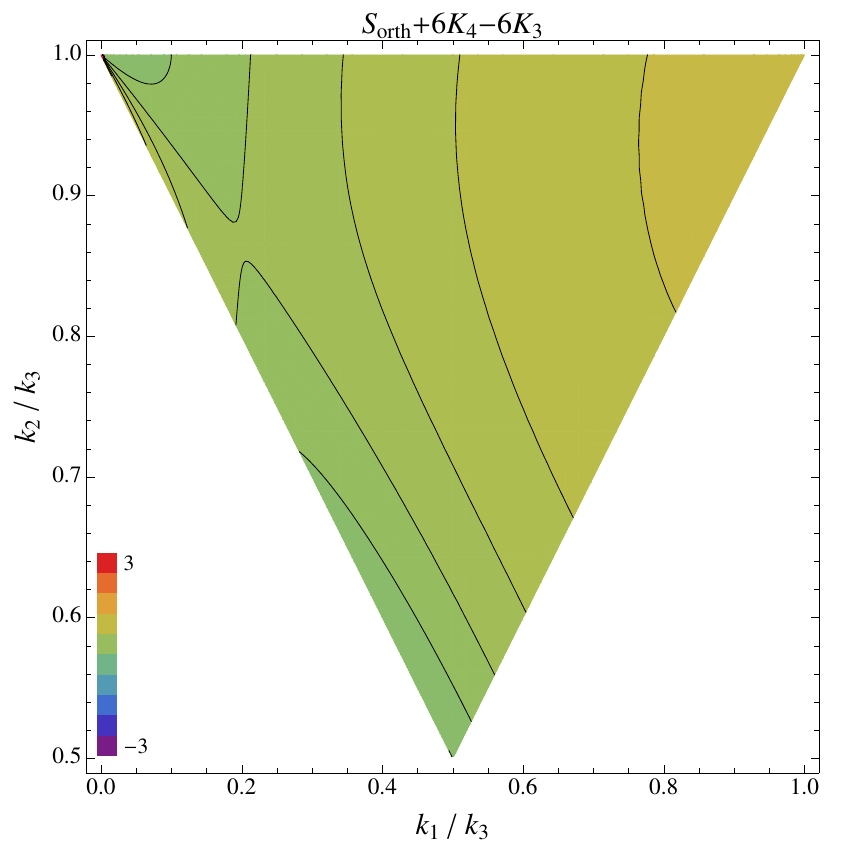}
        }
       \subfloat[subfig5 text][$S_{local}-2{\cal K}_{5}+2{\cal K}_{3}$ ]
        {
        \includegraphics[width=0.33\textwidth]{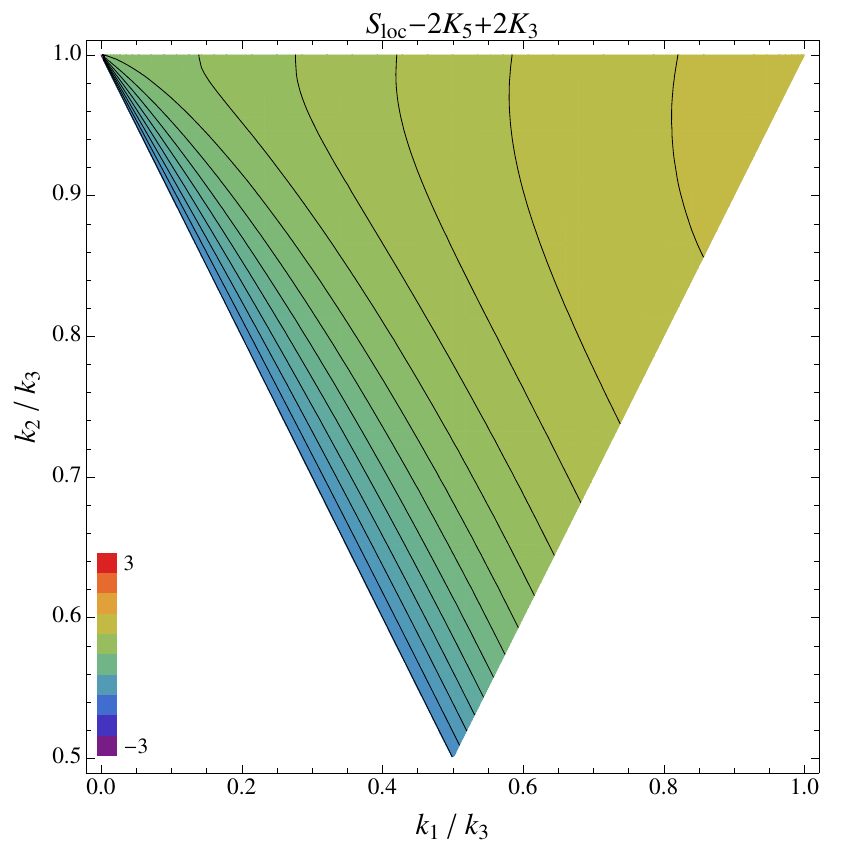}
        }
      \subfloat[subfig6 text][$2{\cal K}_{3}-{\cal K}_{6}$ ]
        {
        \includegraphics[width=0.33\textwidth]{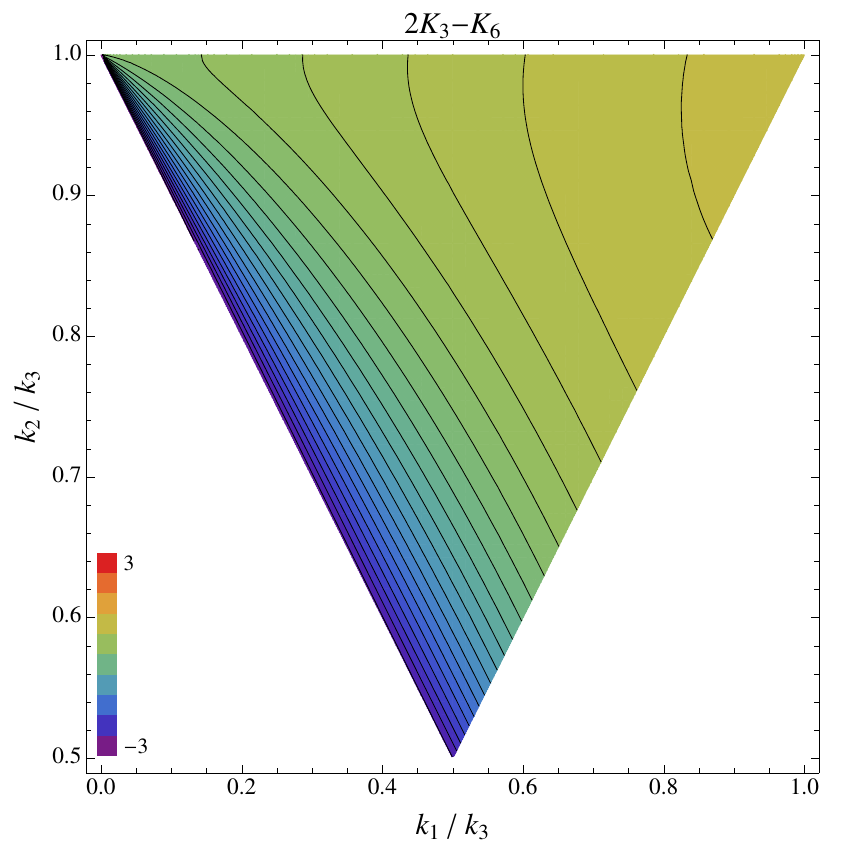}
}
        \caption{Plots showing the comparative spatial distribution of non-Gaussian shape, $S(k_1,k_2,k_3)$, as a function of $k_1/k_3$ and $k_2/k_3$. From left to right we show [top] the local, equilateral, and enfolded separable templates, [middle] the orthogonal(2) template, and non-separable shapes derived from a QSFI model with $\nu=1.3$ and Galileon inflation, and [bottom] shapes contributing to $S_{[-2,1]}$.   All shapes are normalized to unity at the equilateral configuration ($\frac{k_1}{k_3}=\frac{k_2}{k_3}=1$).  The color scales for all but the local and QSFI shapes are the same to aid comparison. \label{fig:triangle-intro} }
       \end{figure}

In Figure \ref{fig:triangle-intro} we show examples of the shapes discussed in the previous section. 
$S_{const}={\mathcal K}_{0}$ is the archetypal component of a family  with similar power over all scales, homogeneous over the whole triangular region plotted.  ``Squeezed'' shapes have a bispectrum amplitude that is peaked in the top left-hand corner of the plot where $\frac{k_1}{k_3}\ll1$ and $\frac{k_2}{k_3}=1$, while ``equilateral'' type shapes peak in the top right-hand corner where $\frac{k_1}{k_3}=\frac{k_2}{k_3}=1$. ``Flattened'' shapes peak along the left edge, where $\frac{k_1}{k_3}+\frac{k_2}{k_3}=1$.
 
Of the shapes we've discussed so far, some clearly fall within these family categories: $S_{local}$, $S_{warm}$, $S_{4}$ and $S_{5}$ are ``squeezed'' shapes, while $S_{equil}$, $S_{DBI}$, $S_{single}$ are ``equilateral'' and $S_{enf}$  is ``flattened''.

There exist other additional  shapes generated by modes ${\cal K}_{3}$ through ${\cal K}_{7}$. For example, Figure \ref{fig:triangle-intro} also includes three shapes  that contribute to $S_{[-2,1]}$ that could describe a general single-field model with Bunch Davies vacuum.  While each vanishes in the squeezed limit by construction, we find they differ from the equilateral shape in still having a component of their signal focused along the flattened configuration. The comparative size of this component correlates with the divergence of the shapes from which they are created, $S_{ortho}$, $S_{local}$, and ${\cal K}_{3}$. 
 
There are shapes that do not fall clearly into any of these families: $S_{ortho(2)}$ peaks in  both the flattened and equilateral configurations, excited states can peak in squeezed and flattened configurations, and  $S_{QSFI}$ shapes interpolate between constant and local properties. Beyond this there are shapes with distinct undulating forms, the $S_{Galileon}$ shape for example, that do not peak at either edges or corners. Moreover, not all shapes within each family are alike. For example, the local and warm shapes both peak in squeezed configurations, but their divergence properties in this region are different, leading to a low cosine between them.

Given the breadth of bispectrum shapes that could be created, and the comparatively loose characteristics on which ``families'' are formed, there is strong motivation to ask how much information we can discern observationally about bispectra. This will help quantify how well we might determine the underlying non-Gaussian shape, if a detection of non-Gaussianity is made.

\section{Forecasting constraints on general shapes}
\label{analysis}

In the following analysis, we apply the separable, divergent basis and template classes from the previous section to assess how we can constrain the shape of primordial bispectra with upcoming CMB data.  Our goal is to quantify what properties of shapes are measurable, and the respective roles of the experimental uncertainties and theoretical priors on determining distinguishability.

Motivated by a broad cross-section of models in the literature, we will focus on shapes described by basis functions $\{\mathcal{K}_{0}-{\cal K}_6\}$ that are nearly scale-invariant and contain terms that are potentially as divergent as $x_{sq}^{-2}$ in the squeezed limit.

We describe the Fisher matrix approach we use assuming a Planck-like CMB experiment in section \ref{sec:fisher}. In section \ref{sec:fisher_results}, we present the results of a principal component analysis for the set of shapes $S_{[-2,r]}$ with different divergence criteria in the squeezed limit imposed.  Doing so generates the experiment's preferred orthogonal basis of best to worst measured bispectrum configurations, the principal components (PCs)  and their corresponding uncertainties, subject to the theoretical prior. We consider the implications for shape normalization and the best-measured $k$-configuration in sections \ref{sec:normalization} and \ref{sec:best_k}, respectively, and finish in section \ref{sec:shape determination} by quantifying our potential ability to determine and distinguish shapes.

\subsection{Fisher matrix approach}
\label{sec:fisher}

We compute the $7\times7$ Fisher matrix for the amplitudes of the basis modes, ${\cal K}_{n}$,  $\{f_{NL}^n, n=0,...,6\}$ defined  in Eq. \eqref{fnln} as
\bea
\label{fisher}
	F(f_{NL}^i,f_{NL}^j) &=& f_{sky} \sum_{abc,pqr} \sum_{\ell_1\ell_2\ell_3} \frac{\partial B^{abc\ (i)}_{\ell_1\ell_2\ell_3}}{\partial f_{NL}^i} (Cov^{-1})^{abc,xyz}_{\ell_1\ell_2\ell_3} \frac{\partial B^{xyz\ (j)}_{\ell_1\ell_2\ell_3}}{\partial f_{NL}^j}, 
\eea
where $\{abc\}$ and $\{xyz\}$ each sum over the 8 possible temperature ($T$) and polarization ($E$) combinations of bispectra: $TTT, TTE, TET, ETT, TEE, ETE, EET, EEE$.  

Given a general primordial shape expanded on the $\{{\cal K}_{n}\}$ basis as in \eqref{fnln}, the corresponding CMB reduced bispectrum is 
\be
b_{\ell_1\ell_2\ell_3}^{abc} = N \sum_{n} f_{NL}^n {\cal K}^{abc (n)}_{l_1l_2l_3},
\ee
where ${\cal K}_{\ell_1\ell_2\ell_3}^{(n)}$ denotes the reduced bispectrum of the basis function ${\cal K}_{n}$ in \eqref{first Kn defined}--\eqref{eq:pprime}.  Here we compute it as
\bea
 {\cal K}^{abc (n)}_{\ell_1\ell_2\ell_3} &=& \frac{1}{N^{(n)}_{perm} k_0^{2(n_s-1)}}\int x^2 dx\left[ {I }^{ap}_{\ell_1}(x) {I }^{br}_{\ell_2}(x) {I }^{cs}_{\ell_3}(x) + \{prs\} \operatorname{ perms}\right]
\\
{I }^{ap}_{\ell}(x) &\equiv& \frac{2}{\pi} \int_{k_{min}}^{k_{max}}  d k k^{p'}\Delta_{\ell}^{a}(k)j_{\ell}(kx)
\eea
where $p'$ is defined as in (\ref{eq:pprime}).

We have modified  the CAMB \footnote{http://camb.info} code \cite{Lewis:1999bs} to numerically evaluate the values of $I_{\ell_1}^{ap}$ and then written code to appropriately combine them to form each $\mathcal{K}_{\ell_1\ell_2\ell_3}^{(n)}$. Specifically, we take $k_{min}=6.6\times10^{-6} \mbox{ Mpc}^{-1}$, $k_{max}=0.56 \mbox{ Mpc}^{-1}$, and $x_{max}=16.5\times10^3 \mbox{ Mpc}$.

We include a note of caution that since the integrals over $k$ and $x$ for cases where $p$ is very negative (positive) depend on having accurate transfer functions at small (large) values of $k$, numerical results for these integrations should be carefully checked for robustness.
To verify the numerical robustness of our results, we have checked that $I^{ap}_{\ell}(x)$ obtained numerically for $p<0$ match the expected analytic result in the Sachs-Wolfe limit.
We have also quantified how the Fisher matrix results quoted in the next section are robust or exhibit instabilities to changes in the accuracy boost parameter in CAMB, which allows for fine resolution in the $k$ and $x$ integrals.
In particular, for the most divergent $\mathcal{K}_6$ mode, which is a combination of the most extreme integrals (with $p=-2$ and 4) and thus we would expect to have the greatest amount of numerical error, we find that the Fisher results quoted in the next section changed by less than $0.01\%$ when the accuracy boost was increased from 1.5 to 2.
However,  we find that the worst measured eigenmode, in the PCA, is far more sensitive to the integral resolution. We find with an  accuracy boost of around 2 we get convergence of a few percent in all but the worst measured mode.  This last mode oscillates with a variation of around 15\% in the standard deviation. This sensitivity in the worst measured mode (which we will find is the least divergent shape in the squeezed limit), can affect the constraints for shapes which have a component described by this mode.  In the following sections, we present our results with these cautions attached when appropriate.

The covariance matrix we use from \cite{Babich:2004yc} is
\bea
 (Cov^{-1})^{abc,xyz}_{\ell_1\ell_2\ell_3} &=& 
   (C^{-1})^{ax}_{\ell_1} \left[ (C^{-1})^{by}_{\ell_2} (C^{-1})^{cz}_{\ell_3}
  + (C^{-1})^{bz}_{\ell_2} (C^{-1})^{cy}_{\ell_3} \right] \nonumber \\
  && +(C^{-1})^{ay}_{\ell_1} \left[ (C^{-1})^{bz}_{\ell_2} (C^{-1})^{cx}_{\ell_3} 
  +  (C^{-1})^{bx}_{\ell_2} (C^{-1})^{cz}_{\ell_3} \right] \nonumber \\
  && +(C^{-1})^{az}_{\ell_1} \left[ (C^{-1})^{bx}_{\ell_2} (C^{-1})^{cy}_{\ell_3}
  + (C^{-1})^{by}_{\ell_2} (C^{-1})^{cx}_{\ell_3} \right],
\eea
with
\bea
 (C^{-1})^{ax}_{\ell} &=& \left( \begin{array}{cc} \hat{C}_\ell^{TT} & \hat{C}_\ell^{TE} \\ \hat{C}_\ell^{TE} & \hat{C}_\ell^{EE} \end{array} \right)^{-1}
 \\
\hat{C}_\ell^{ax} &=& C_\ell^{ax} + N_{\ell}^{ax}.
\eea
Here $f_{sky}$ is the overall fraction of the sky observed, and we assume $f_{sky}=0.8$. $N_{\ell}^{ax}$ is the instrument noise for a correlation between observables $a$ and $x$. We model CMB noise by considering the three lowest frequency bands of the Planck HFI instrument for temperature and E-mode polarization, as described in the Planck Bluebook \cite{Planck_bluebook}. We assume each frequency channel has Gaussian beam profile of width $\theta_{FWHM}$ and isotropic noise with error in $X = T,E$ of $\sigma_{X}$. The noise in each frequency channel $c$ is then given by
\bea
N_{\ell,c}^{ax} &=& (\sigma_{x,c}\theta_{fwhm})^2e^{\ell(\ell+1)\theta^{2}_{fwhm,c}/8\ln 2}\delta_{ax}
\\
N_{\ell}^{ax} &=& \left[\sum_c\left(N_{\ell,c}^{ax}\right)^{-1}\right]^{-1}.
\eea

Our fiducial flat $\Lambda$CDM cosmology is described by the following parameters, which are consistent with the latest WMAP 9-year constraints \cite{Hinshaw:2012fq}: $\Omega_bh^2=0.02258$, $\Omega_ch^2=0.1109$, $\Delta^2_\mathcal{R}(k_0)=2.43\times10^{-9}$, $n_s=0.963$, and $\tau=0.088$.  As has been done in other recent Fisher forecasts on non-Gaussianity parameters, such as \cite{Sefusatti:2009xu}, we consider the uncertainties on the non-Gaussian amplitudes independent of the uncertainties in the fundamental cosmological parameters that also affect the power spectrum, as these are comparatively small relative to the uncertainties from the bispectrum shape functions \cite{Liguori:2008vf}.  For this initial analysis, we  neglect  the effect of imperfect measurements of the lensing signal \cite{Lewis:2011fk,Pearson:2012ba}, secondary anisotropies \cite{Komatsu:2001rj}, and inhomogeneous sky coverage/noise on the constraints (e.g. \cite{Smith:2006ud,Fergusson:2009nv}).

\subsection{Fisher matrix results}
\label{sec:fisher_results}

A general bispectrum can be expanded in terms of either ${\cal K}_n$ or the component shapes, $\{S_X\}$, in $S_{[R,r]}$, given in (\ref{eq:Sm10})-(\ref{eq:Sm2m2}), 
\bea
	\frac{B_\Phi(k_1,k_2,k_3) (k_1k_2k_3)^2}{N} &=& f_{NL}S=  \sum_n f_{NL}^n {\cal K}_{n}(k_1,k_2,k_3)  =  \sum_X f_{NL}^X S_{X}(k_1,k_2,k_3) \ \ \ 
\eea
While the Fisher matrix we used based on $S_{[R,r]}$ automatically includes the additional priors to constrain the divergence properties, these could also be introduced  into the ${\cal K}_n$ Fisher analysis by using Lagrange multipliers to systematically impose each divergence constraint. The latter makes no assumption a priori about what linear combinations of the shapes given in Table \ref{tab:subshapes} should have their amplitudes constrained. While we use the shape expansion in our discussion below, we investigated both approaches and found they led to consistent conclusions. 

We use the Fisher matrix in terms of ${\cal K}_n$ to construct Fisher matrices for the component shapes in $S_{[-2,r]}$ for $r=-2,-1,0,1$. In Table \ref{tab:correll} we give the $\ell$-space correlation coefficients based on (\ref{eq:ellcorr}), but here weighted by the data covariance between pairs of the component shapes, $S_X$ and $S_Y$,
\bea
Corr_{\ell}(S_{X},S_{Y}) &=& \frac{F_{XY}}{\sqrt{F_{XX}F_{YY}}}.
\eea
This gives a measure of the similarity of the component shapes based on how they are measured by the survey, integrated over all $\ell$ combinations. 

We find the similarity between pairs of the four basis shapes in $S_{[-2,1]}$, each of which vanishes in the squeezed limit, are primarily related to the divergence of the shapes  from which they are derived. $S_{equil}$ and $S_{ortho}+6{\cal K}_4-6{\cal K}_3$  are very similar to each other, while $S_{local}-2{\cal K}_5+2{\cal K}_3$ and $2{\cal K}_3-{\cal K}_6$ also have a high degree of overlap. Interestingly the $S_{local}-2{\cal K}_5+2{\cal K}_3$ and $2{\cal K}_3-{\cal K}_6$ shapes also have significant similarities with the shapes that diverge as $x_{sq}^{0}$. This is derived from their strong signal along the configurations between squeezed and flattened configurations, as discussed in section \ref{sec:classification}. The shape with $x_{sq}^{-1}$ divergence  constructed from the $R=-2$ modes, $2{\cal K}_5-{\cal K}_6$, is highly degenerate with the local template; essentially this implies the two are indistinguishable from one another using the CMB data.

\begin{table}
\begin{tabular}{|c|l|| c|c|c|c||c|c||c|c||c|}
\hline
\multicolumn{2}{|r||}{Divergence ,    ${x_{sq}}^{n}$, $n=$ \ \ \ } & \multicolumn{4}{c||}{1} &  \multicolumn{2}{|c||}{0} & \multicolumn{2}{|c||}{-1} & -2 \\ 
\hline
 $n$  &$Corr_{\ell}$ & 
\begin{sideways} S$_{equil}$  \end{sideways} & 
\begin{sideways}  $S_{ortho}+6{\cal K}_4-6{\cal K}_3$ \ \ \ \end{sideways} &
\begin{sideways} $S_{local}-2{\cal K}_5+2{\cal K}_3$ \end{sideways}&
\begin{sideways} $2{\cal K}_3-{\cal K}_6$ \end{sideways} &
\begin{sideways} $S_{ortho}$   \end{sideways} &
\begin{sideways} $2{\cal K}_4-{\cal K}_6$ \end{sideways} & 
\begin{sideways} $S_{local}$   \end{sideways}& 
\begin{sideways} $2{\cal K}_5-{\cal K}_6$ \end{sideways}& 
\begin{sideways} ${\cal K}_6$ \end{sideways}
   \\ \hline \hline
 \multirow{4}{*}{1}&   S$_{equil}$   & 1 & 0.97 & 0.41 & 0.07 & -0.11 & 0.33 & 0.23 & 0.21 & 0.003  \\ 
  & S$_{ortho}+6{\cal K}_4-6{\cal K}_3$ \ \  &   & 1 & 0.24&-0.10 & -& -& -& -&-
 \\ 
 &$S_{local}-2{\cal K}_5+2{\cal K}_3$ &  &  & 1& 0.94 &0.78&0.98&- &- &-
   \\
 &$2{\cal K}_3-{\cal K}_6$  &  & & & 1 &0.80& 0.95& -0.29&-0.20  &-0.03
 \\ \hline
 \multirow{2}{*}{0} & $S_{ortho}$&  & & & & 1& 0.72&-0.48 &-0.40 &-0.06
 \\ 
 &$2{\cal K}_4-{\cal K}_6$  &  & & & & &1 &-0.12 & -0.04&-0.01
 \\  \hline
 \multirow{2}{*}{-1}  &$S_{local}$   &  & & & & & &1 & 0.99&0.66
 \\ 
 &$2{\cal K}_5-{\cal K}_6$ & &  & & & & & &1 &0.68
 \\ \hline
\end{tabular}
\caption{Correlation coefficients between shapes that diverge as $x_{sq}^{n}$. These shapes are components in the general template classes, $S_{[-2,r]}$, for $r\le n$.}\label{tab:correll}
\end{table}

The unmarginalized errors, $\sigma(f_{NL}^X)=1/\sqrt{F_{XX}}$, give the uncertainty in the measurement of a specific template if the underlying theory is known to be wholly described by that template. We find these are comparatively insensitive to the integral resolution discussed in section \ref{sec:fisher}.  The covariance matrices obtained from inverting the Fisher matrices give the uncertainties on the amplitudes of the component shapes, $\sigma(f_{NL})$, marginalized over the freedom allowed by each model. The marginalization does make the results precision dependent in the worst measured mode, i.e. the results are accurate to better than 15\%.

We summarize the results in Table \ref{tab:sigmafNL}. 
The covariance matrix in each case can be diagonalized to obtain the orthonormal eigenvectors,
\bea
\hat{e}_i &=& \sum_{X} c_{iX}S_{X},
\eea
and associated eigenvalues, which give the variances $\sigma^2(b_i)$ in the amplitudes of the eigenvectors.  These then provide a way to rank the best to worst measured bispectra.  Given this orthonormal basis, any general bispectrum may be expanded as
\be
f_{NL}S = \sum_i b_i \hat{e}_i.
\ee
The principal components obtained by diagonalizing the covariance matrix are not immediately `shapes' in the way we considered so far. They  have unit norm  with respect to the component shape basis, $\sum_{X}|c_{iX}|^2=1$, rather than being normalized at the equilateral configuration, $\sum_{X} c_{iX}S_{X}(k_0,k_0,k_0)=1$.

If we restrict the shapes to those described by the first three modes, marginalization does not significantly alter the constraints from the unmarginalized errors, i.e. the three common templates are essentially the principal components (PC) of the covariance matrix, with the eigenvalues showing that the more divergent the shape, the better it is measured.  
In contrast, when extended to general shapes, constructed of all seven modes, we find marginalized errors for individual shapes are far larger because of observational similarities between shapes of similar divergence, or similar properties in the flattened limit. It seems that only $K_6$ is well constrained if any shape from the $S_{[-2,r]}$ type is allowed.

When extended to shapes constructed of seven modes, the correspondence between the PC's and  divergence remains. We find that, in general, divergence in the squeezed limit, followed by a second divergence measure, corresponding to the signal near the flattened configurations, can be used as coarse indicators  of comparative constraining power with the CMB.  For the general shape without any additional divergence constraints, the best measured PC is almost completely composed of the most divergent shape, ${\cal K}_{6}$. The second best measured PC has dominant contributions from $S_{local}$ and $2{\cal K}_{5}-{\cal K}_{6}$ with which it is very degenerate. 
If the general shape is restricted to have vanishing divergence in the squeezed limit, then the best measured PC is very similar to a shape like $2{\cal K}_3-{\cal K}_{6}$ which has large signal in the flattened configurations despite vanishing in the squeezed limit.  The next best measured PC is then similar to shapes like equilateral or the orthogonal-derived shape $S_{ortho}+6{\cal K}_{4}-6{\cal K}_{3}$, which has less power on flattened configurations.
In both cases, none of the templates look like the two worst measured modes, which exhibit large oscillatory features along flattened configurations.

\begin{table}
\centering
\begin{tabular}{|c|l||c||c|c||c|c|c|c|}
\hline
 & & &  \multicolumn{6}{|c|}{ $\sigma(f_{NL})$ marg.$^d$ over shape }
\\  \cline{4-9} 
 Divergence &   \multirow{2}{*}{ Shape} & Unmarg. & \multicolumn{2}{|c||}{  $S_{[-1,r]} $} & \multicolumn{4}{|c|}{  $S_{[-2,r]} $}
 \\
  \cline{4-9} 
   ${x_{sq}}^{n}$ && $\sigma(f_{NL})$&  $r$=0  & -1 &$r$=1& 0  & -1  & -2 
 \\ \hline \hline
\multirow{4}{*}{1} & S$_{equil}$  & 43 & 44 & 45 & 351 & 365 & 387 & 404
\\
& S$_{ortho}+6{\cal K}_4-6{\cal K}_3$ & 41   & - & - & 293 & - &-  & -
\\
& $S_{local}-2{\cal K}_5+2{\cal K}_3$ &32 & - & -  & 920 &  1064 & - & - 
 \\
&  $2{\cal K}_3-{\cal K}_6$  & 18 & - & - & 468 & 742 & 1425 & 1428
\\ \hline
\multirow{2}{*}{0} & $S_{ortho}$   & 19  & 19 & 22 &- & 362 &  364 & 366
\\
& $2{\cal K}_4-{\cal K}_6$  & 23 & - & -& - & 1000 & 1018 & 1034
\\ \hline
\multirow{2}{*}{-1} & $S_{local}$   & 3 & - & 4 & - & - & 1073 & 1081
\\
 & $2{\cal K}_5-{\cal K}_6$  & 4& - & -& - & - & 1074 & 1082
\\ \hline
\multirow{1}{*}{-2} & ${\cal K}_6$  &  0.011 & - & -& - &- &-  & 0.018
\\ \hline
\end{tabular}
\caption{The uncertainties on the amplitudes of the component shapes, in the general template classes $S_{[-2,r]}$, that  diverge as $x_{sq}^{r}$ in the squeezed limit. We give both the unmarginalized errors, assuming the underlying shape is exactly described by the component shape, and the marginalized errors if we allow the shape to be a general linear combinations of components consistent with the prior on the divergence properties.
}
\label{tab:sigmafNL}
\end{table}
\begin{table}[t!]

\begin{center}
\begin{tabular}{|c||c|c||c|c||c|c||c|c||}
\hline
Shape & \multicolumn{2}{|c||}{$S_{[-2,-2]}$} &  \multicolumn{2}{|c||}{$S_{[-2,-1]}$} &  \multicolumn{2}{|c||}{$S_{[-2,0]}$} &  \multicolumn{2}{|c||}{$S_{[-2,1]}$}
\\ \hline
  &  $\sigma(b_i)$ & $\sigma(f_{NL,i})$ &  $\sigma(b_i)$ & $\sigma(f_{NL,i})$ &   $\sigma(b_i)$ & $\sigma(f_{NL,i})$&  $\sigma(b_i)$ & $\sigma(f_{NL,i})$\\ \hline
$\hat{e}_1$ &0.0084& 0.016	&  2.8 &  4		& 20 	&  18  & 31    & 11 \\
$\hat{e}_2$ &  4 	&  5	&  22 	& 18 	& 38 	&  32  & 138   & 45 \\
$\hat{e}_3$ &  24 	&  18 	&  38 	& 32 	& 49 	& 39   & 1321  & 26 \\
$\hat{e}_4$ &  38 	&  32	&  491	& 41 	& 1321 	& 25   & 7518  & 28 \\
$\hat{e}_5$ & 522 	&  43 	& 1420	& 22  	& 9576	& 16   &  	    &  	 \\
$\hat{e}_6$ &1505 	&  19 	& 9576	& 16 	& 		& 	   & 	    &	 \\
$\hat{e}_7$ &9576 	&  16	& 		& 		& 		& 	   & 	    &
\\\hline
\end{tabular}
\end{center}
\caption{Properties of the principal components for each template class $S_{[-2,r]}$ in terms of their component shapes. The properties in the squeezed limit is determined by the value of $r$. The table provides uncertainties, for a unit norm eigenvector, $\sigma(b_i)$, and an effective $\sigma(f_{NL}(\hat{e}_i)) $, when the eigenvector is normalized consistently at the equilateral configuration.}\label{table:pca_te}
\end{table}%
\begin{figure}[t!]
    \centering
    \subfloat[subfig1 text][$\hat{e}_1$]
        {
        \includegraphics[width=0.33\textwidth]{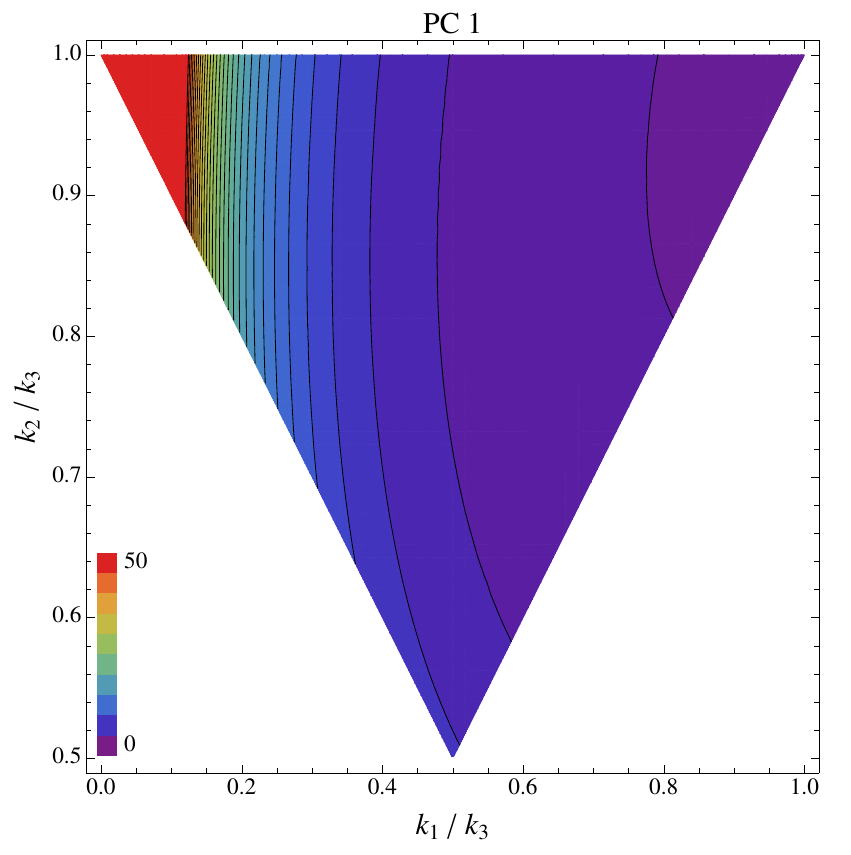}
        \label{fig:PC1_TE}
        }
    \subfloat[subfig1 text][$\hat{e}_2$]
        {
        \includegraphics[width=0.33\textwidth]{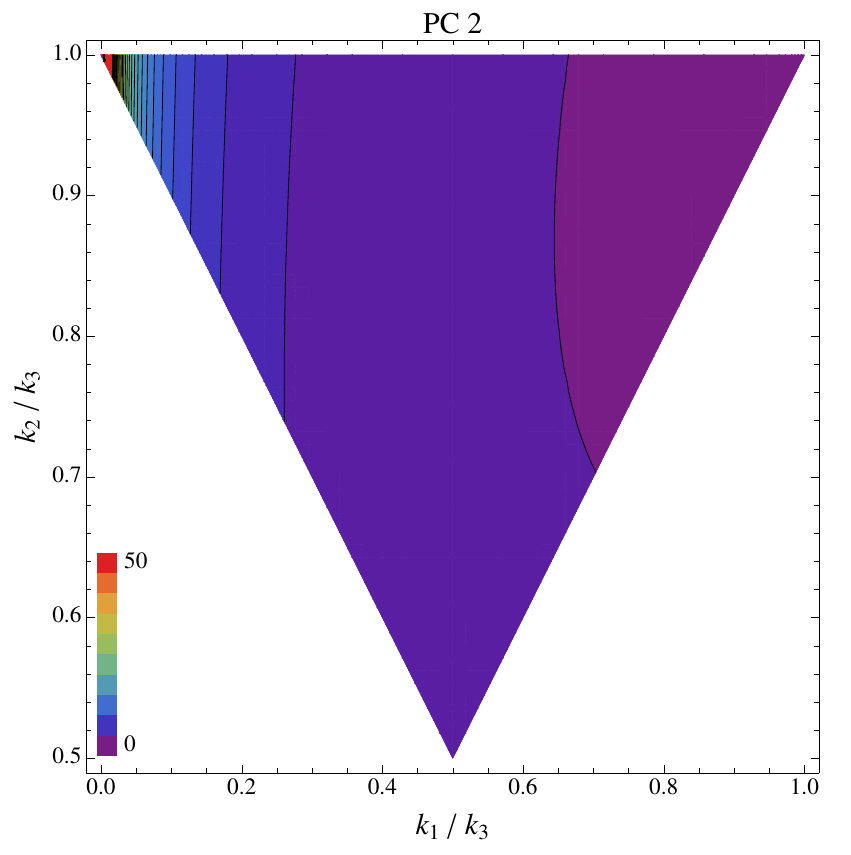}
        \label{fig:PC2_TE}
        }
    \subfloat[subfig3 text][$\hat{e}_3$]
        {
       \includegraphics[width=0.33\textwidth]{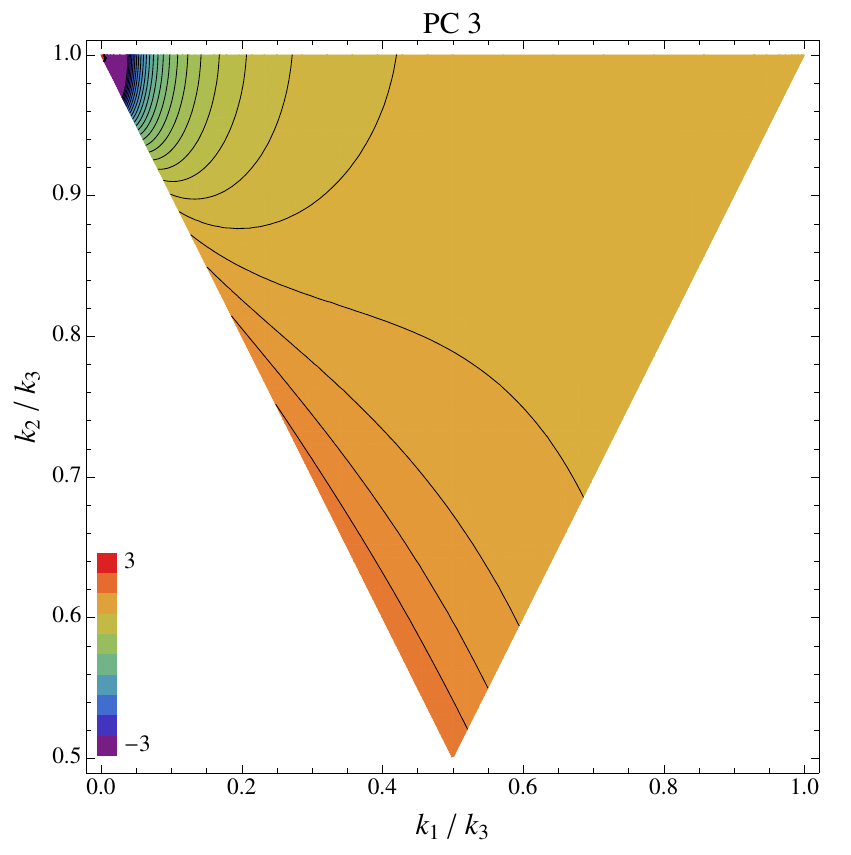}
        \label{fig:PC3_TE}
        }
            \\
            \subfloat[subfig4 text][$\hat{e}_4$]
        {
        \includegraphics[width=0.33\textwidth]{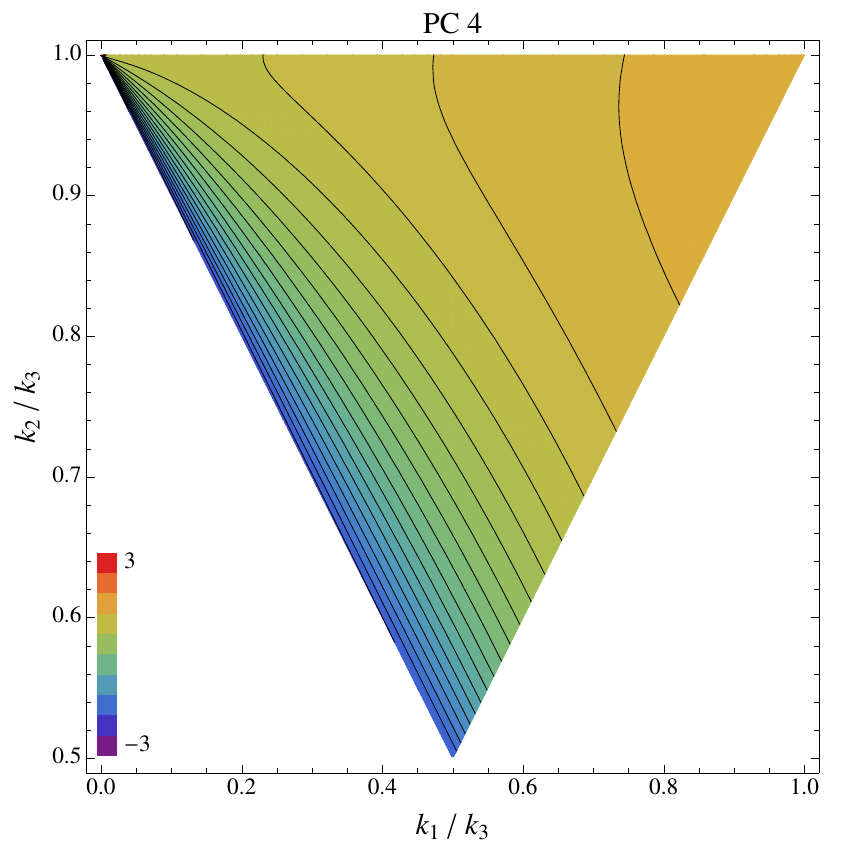}
        \label{fig:PC4_TE}
        }
    \subfloat[subfig3 text][$\hat{e}_5$]
        {
        \includegraphics[width=0.33\textwidth]{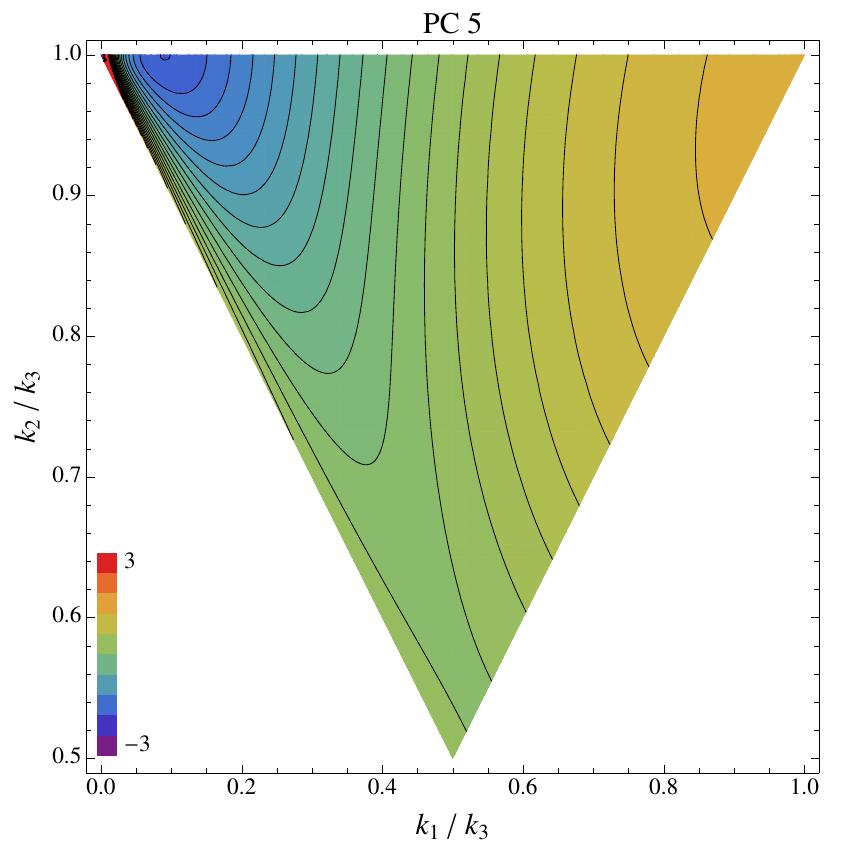}
        \label{fig:PC5_TE}
        }
    \subfloat[subfig4 text][$\hat{e}_6$]
        {
        \includegraphics[width=0.33\textwidth]{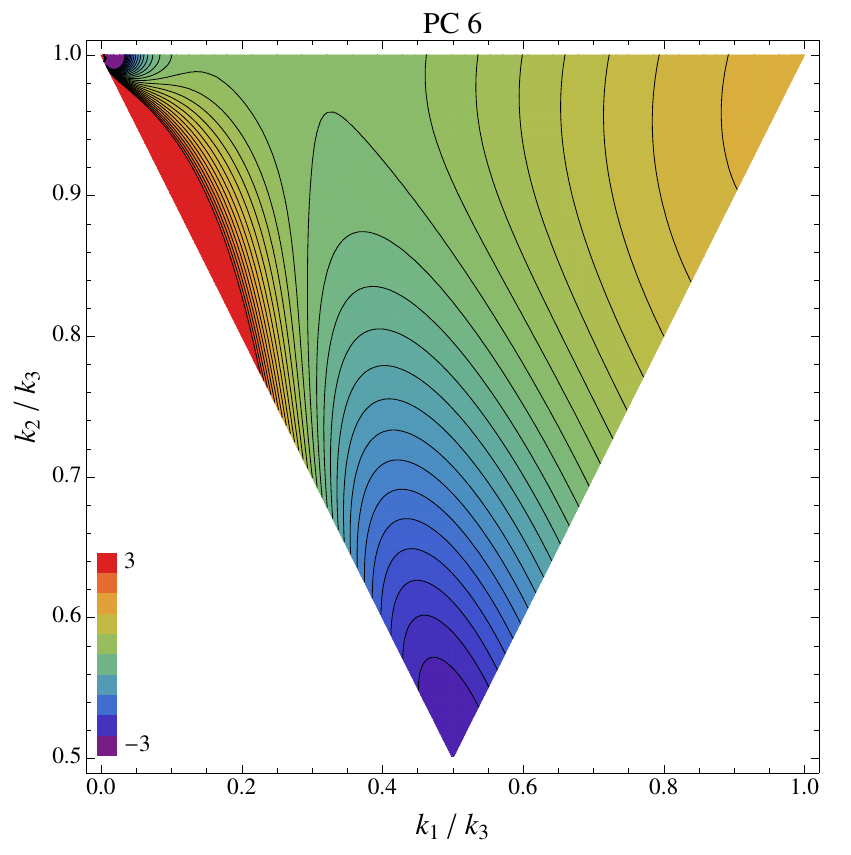}
        \label{fig:PC6_TE}
        }
            \\
    \subfloat[subfig3 text][$\hat{e}_7$]
        {
        \includegraphics[width=0.33\textwidth]{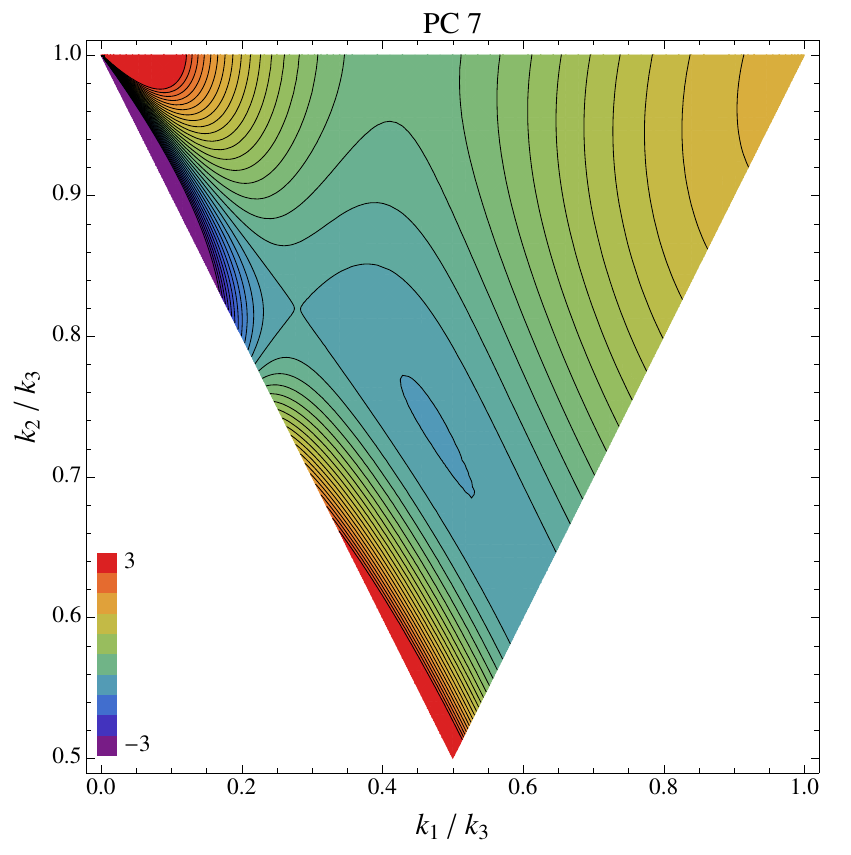}
        \label{fig:PC7_TE}
        }
    \caption{Configurations of the principal components for $S_{[-2,-2]}$,  a general shape that can be as divergent as $x_{sq}^{-2}$ in the squeezed limit. The plots show the amplitude of the eigenvectors for the best $\hat{e}_1$ to worst $\hat{e}_7$ measured modes as a function of $\frac{k_1}{k_3}$ versus $\frac{k_2}{k_3}$. The principal components are each normalized to be unity at the equilateral configuration. 
    }
    \label{fig:PC_TE}
\end{figure}

\subsection{Drawbacks of normalization at the equilateral configuration}
\label{sec:normalization}

As stated earlier, the PC's as they are originally generated, are not shapes in the usual sense because they are not bispectra normalized at $k_1=k_2=k_3=k_0$.   They have a unit norm in terms of the basis shapes. With this normalization, as usual in PCA, their eigenvalues quantify which  combinations  of the basis shapes  are best and worst constrained by data, and the eigenvectors can be combined to create general shapes.  

We can convert $\sigma(b_i)$ to an effective $\sigma(f_{NL}(\hat{e}_i))$, corresponding to the amplitude of each eigenvector shape normalized in the conventional way,  $\sigma(f_{NL}(\hat{e}_i)) =  |\sigma(b_i)\hat{e}_i(k_0,k_0,k_0)|$.   Table \ref{table:pca_te} gives the values of $\sigma(b_i)$ and $\sigma(f_{NL}(\hat{e}_i))$. We quote the results when both temperature and polarization data are included. We find that the exclusion of the E-mode polarization from the Fisher analysis does not noticeably change the shape of the principal components, but does increase the eigenvalues by about a factor of $\sim 1-3$ across all eigenvectors.
The constraints on all but the last eigenvalue under each divergence constraint shown in Table \ref{table:pca_te} are accurate to a few percent. The worst measured eigenmode is measured to $\sim 15\%$ accuracy.

Normalizing our PC's at the arbitrarily chosen equilateral configuration allows us to compare them to other shapes consistently at one point in $k$-space.  $\sigma(f_{NL})$ does not in itself, however, quantify a shape's overall variance across all $k$. An analogous situation arises in quoting uncertainties on the power spectrum amplitude from two different surveys, say a large-scale CMB survey and a galaxy survey.  Both surveys could quote uncertainties at a common arbitrary scale, say $k_0=0.05h/Mpc$, but while this uncertainty might represent the best measured scale for the galaxy survey, it would grossly overstate the minimum uncertainty in the CMB survey, which is best measured at a much larger scale.

It is entirely possible for a well measured mode to have a significant part of its small variance located in the equilateral configuration, while a poorly measured mode could have its lowest  variance in the equilateral configuration but be poorly measured over other regions of $k$-space. Indeed we find this to be the case, given that the best measured shapes have signal peaked near the squeezed, rather than equilateral, configuration. This means that $\sigma(f_{NL})$ is not a useful measure in itself to assess how well a shape can be measured. This shortcoming of the conventional normalization has been discussed previously in other studies, e.g. \cite{Creminelli:2005hu} and \cite{Fergusson:2009nv}, where alternative normalization schemes based on an integrated total amount of non-Gaussianity have been proposed.

The overall spread in uncertainties from the best to worst eigenvector is much reduced when normalized at the equilateral configuration and can in some cases produce a switch in the ordering of the modes for $\sigma(f_{NL})$ relative to that of $\sigma(b_i)$.  This does not present an inconsistency in the analysis, but simply demonstrates the perils of considering a normalization at an arbitrary scale.

Figure \ref{fig:PC_TE} shows the variety of profiles in the 2-dimensional ($\frac{k_1}{k_3}, \frac{k_2}{k_3}$) space shown in the triangle plots. Given that the power spectrum we consider is not perfectly scale invariant, there is some small dependency of the bispectrum amplitude on the value of $k_3$, described by $p'$ in (\ref{eq:pprime}). The spatial profiles, however, in terms of $\frac{k_1}{k_3}$ and $\frac{k_2}{k_3}$  are $k_3$-independent.  

The  gradients in the PC configurations reflect the rough ordering from squeezed to flattened to equilateral as the modes span from best to worst. The  complementarity of the eigenvectors, reflected by the different directions of gradients of the signals in the configuration space, has implications for the location of the best measured configuration, as we discuss in section \ref{sec:best_k}.

\begin{figure}[t]
    \centering
    \subfloat[subfig1 text][$\hat{e}_1$]
        {
        \includegraphics[width=0.5\textwidth]{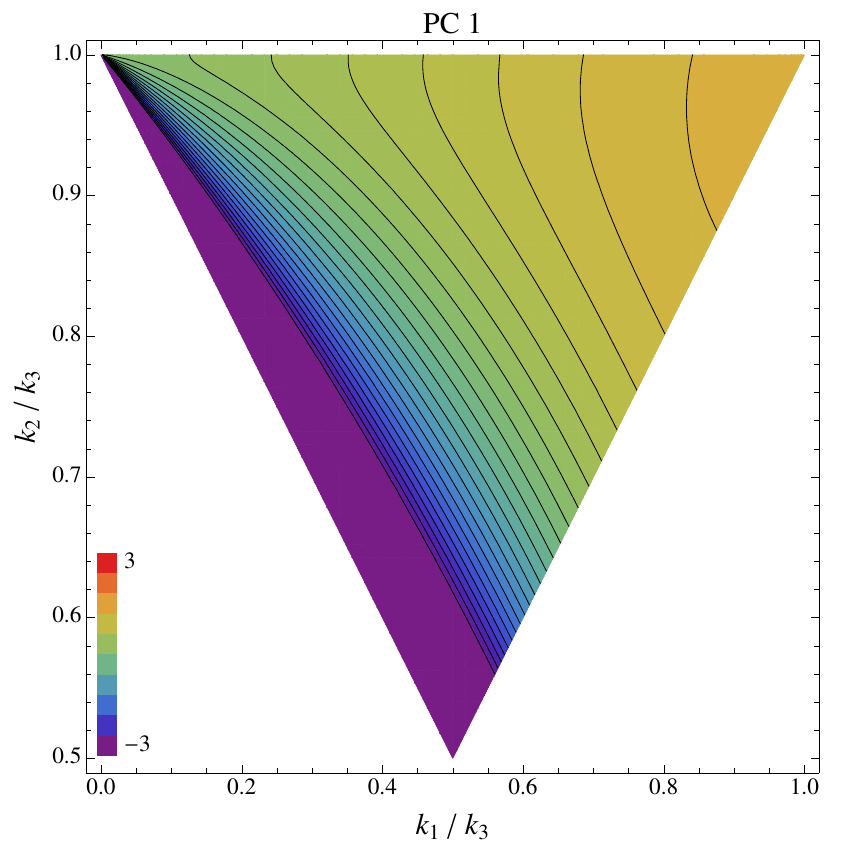}
        \label{fig:zeroDiv_PC1_TE}
        }
    \subfloat[subfig1 text][$\hat{e}_2$]
        {
        \includegraphics[width=0.5\textwidth]{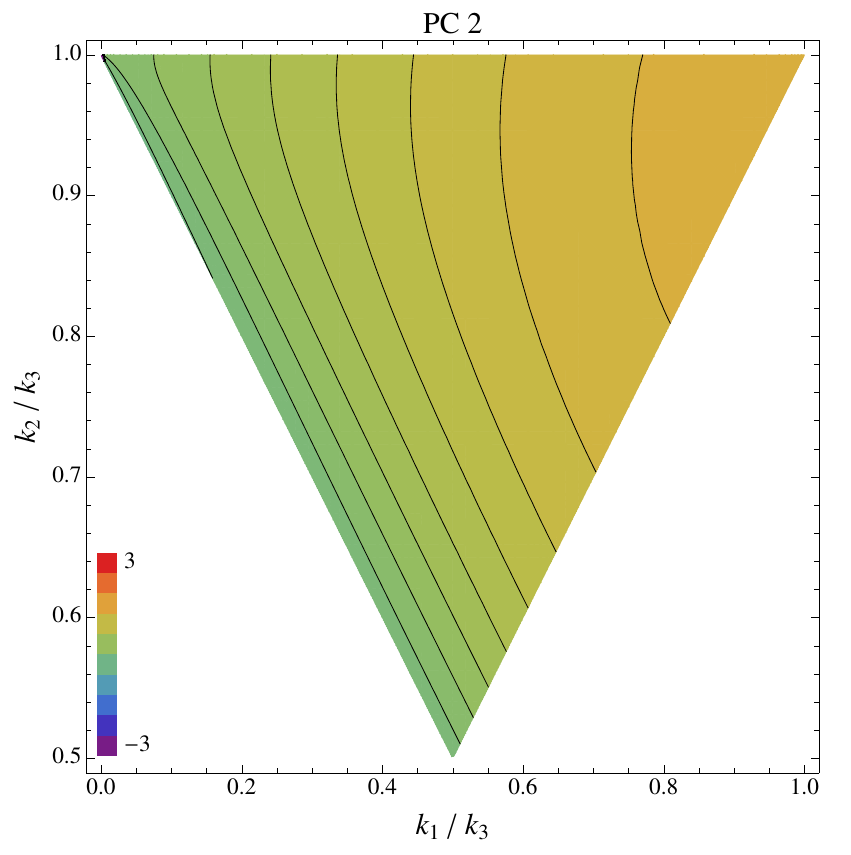}
        \label{fig:zeroDiv_PC2_TE}
        }
        \\
    \subfloat[subfig3 text][$\hat{e}_3$]
        {
        \includegraphics[width=0.5\textwidth]{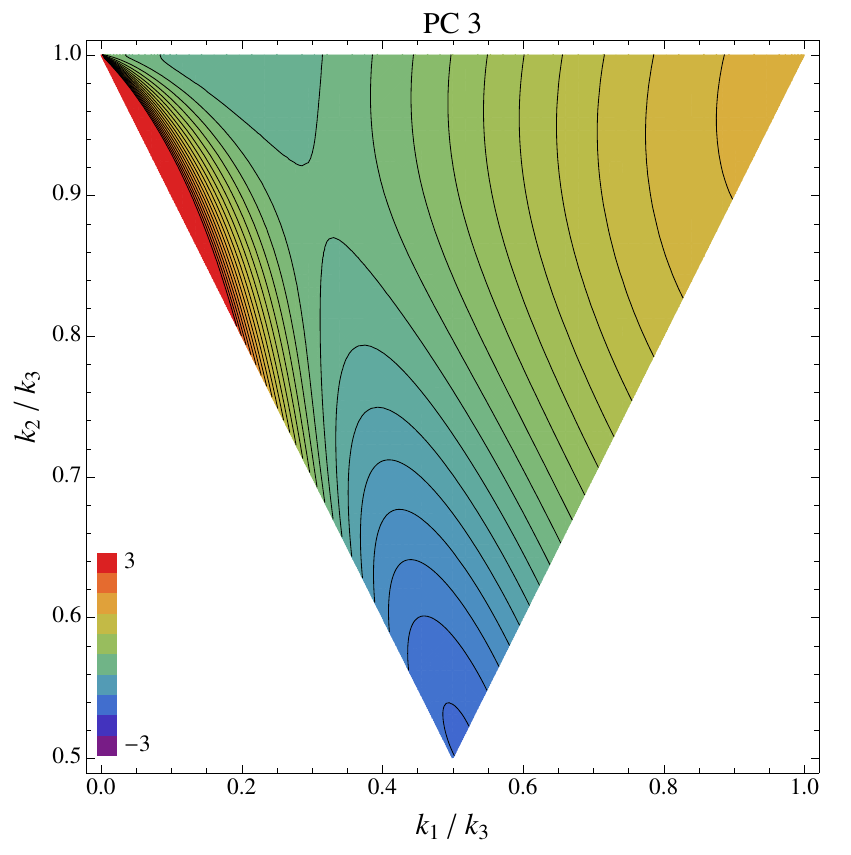}
        \label{fig:zeroDiv_PC3_TE}
        }
    \subfloat[subfig4 text][$\hat{e}_4$]
        {
        \includegraphics[width=0.5\textwidth]{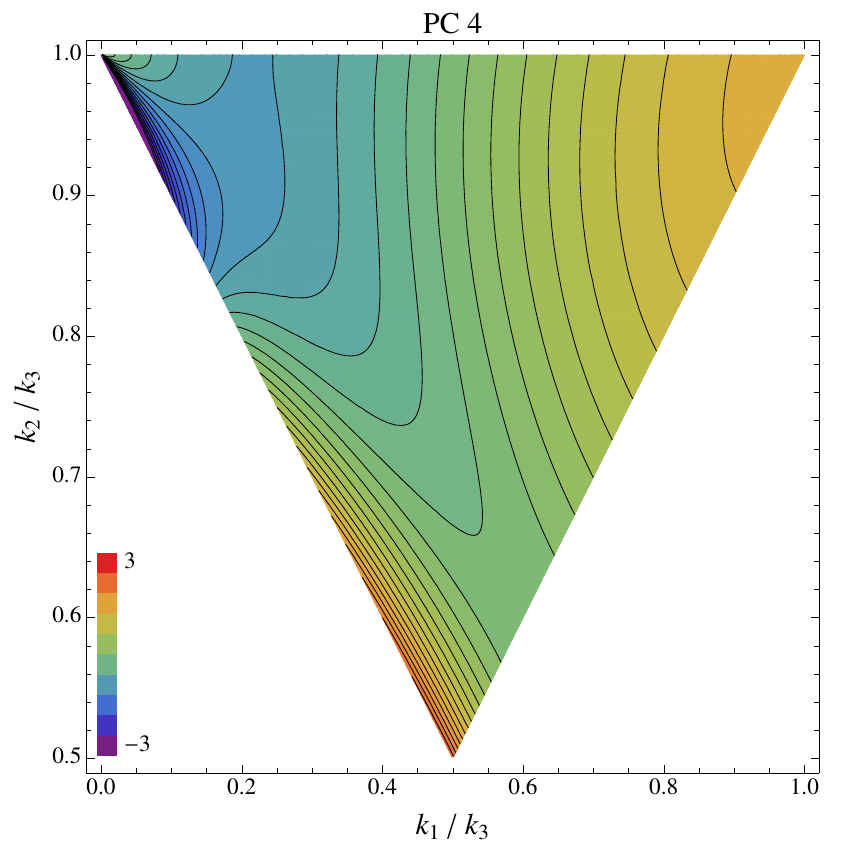}
        \label{fig:zeroDiv_PC4_TE}
        }
    \caption{As in Figure \ref{fig:PC_TE}, but showing the configurations of the principal components for $S_{[-2,1]}$,  a general shape that vanishes in the squeezed limit.}
    \label{fig:zeroDiv_PC_TE}
\end{figure}

\subsection{Best measured $k$-configurations}
\label{sec:best_k}

In the analysis that follows, we  avoid splitting up bispectra into shapes and amplitudes,  normalized at an arbitrary configuration. Instead we consider the overall constraints on the bispectrum, $B(k_1,k_2,k_3)$, itself up to the constant normalization, given in (\ref{eq:BPhi}), $f_{NL}S=k_1^2k_2^2k_3^2 B(k_1,k_2,k_3)/N$. 

The eigenmodes and eigenvalues from the PCA provide a way to compute an error on a general $k$-space bispectrum. We can calculate the posterior distribution of the uncertainties on $f_{NL}S$ given the data, $D$, with a theoretical prior given by the eigenvectors $\{\hat{e}_i\}$,
\bea
p(f_{NL}S|D) &=& \int \prod_{i=0}^{n}db_i p(f_{NL}S|b_i) p(b_i|D),
\\
p(f_{NL}S|b_i) &=& \delta(f_{NL}S - \sum_{i=0}^{n} b_i\hat{e}_i(k_1,k_2,k_3)),
\\
 p(b_i|D)&=&\frac{1}{\sqrt{2\pi}\sigma(b_i)}\exp\left(-\frac{b_i^2}{2\sigma^2(b_i)} \right).
\eea
Under this assumption of Gaussian errors this gives the commonly used result,
\be
\sigma^2 \left( f_{NL}S (k_1,k_2,k_3)\right) = \sum_i \sigma^2(b_i) \hat{e}_i(k_1,k_2,k_3)^2.
\ee
This equation for computing the error can be applied to each set of PC's generated for each divergence scenario in the previous section. The errors in the ($\frac{k_1}{k_3},\frac{k_2}{k_3}$) configuration space can  be plotted and  the best measured $k$-configuration, and the associated uncertainty, calculated for each scenario.

$\sigma(f_{NL}S)$ varies only very weakly across  slices in $k_3$; its functional form can be divided into a dependence on ($\frac{k_1}{k_3}, \frac{k_2}{k_3}$) and a weak dependence on $k_3$,  going as $k_3^{2(n_s-1)}$, for fixed $\frac{k_1}{k_3}$ and $\frac{k_2}{k_3}$. For our choice of theoretical priors on the model, $\sigma(f_{NL}S)$ decreases with increasing $k_3$. This is because  the noise scales as the signal for the near scale-invariant theoretical prior we impose.  An alternative prior would give very different dependencies on  $k_3$. For example if we were to remove the theoretical prior all together and model the bispectrum amplitude as bins in $k$, the only constraints on the model come from the observational uncertainties, and the noise would diverge exponentially on small scales.  

\begin{figure}[t]
    \centering
    \subfloat[subfig1 text][Diverges as $1/k^2$ in the squeezed limit.]
        {
\includegraphics[width=0.5\textwidth]{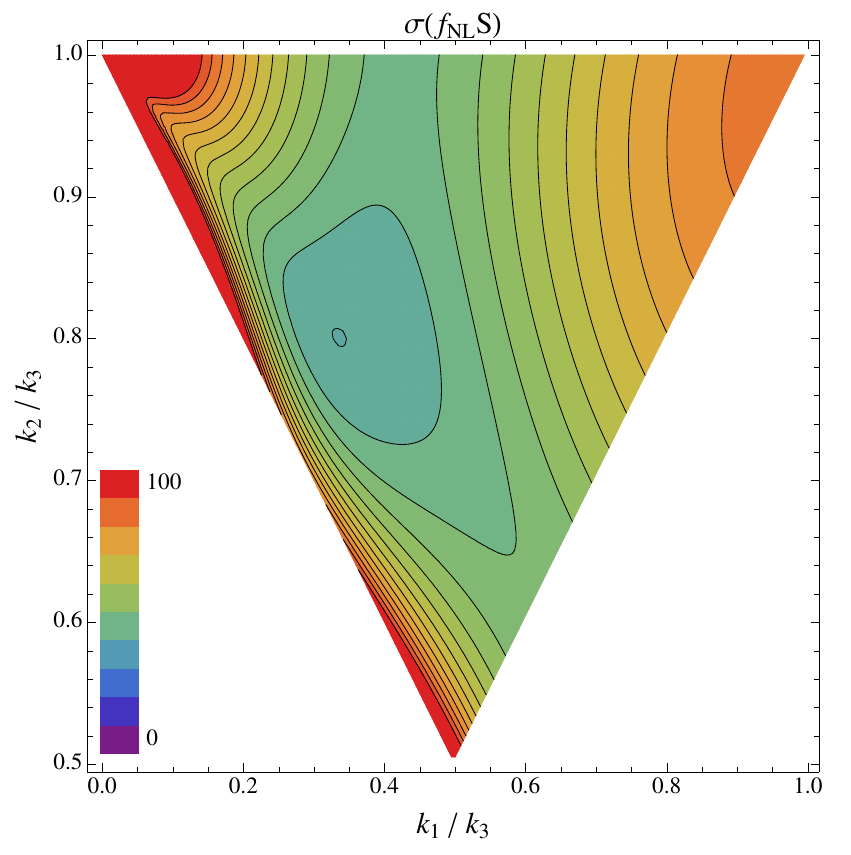}
        }
    \subfloat[subfig4 text][Vanishing divergence in the squeezed limit.]
        {
\includegraphics[width=0.5\textwidth]{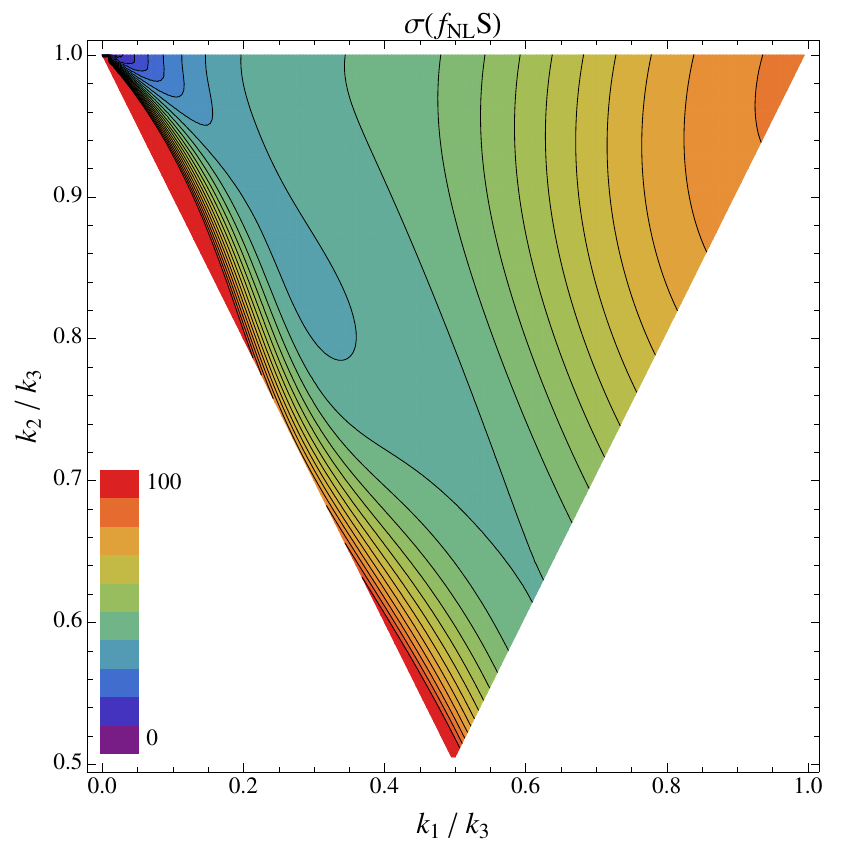}
        }
        \\
    \subfloat[subfig4 text][Diverges as $1/k$ in the squeezed limit.]
        {
\includegraphics[width=0.5\textwidth]{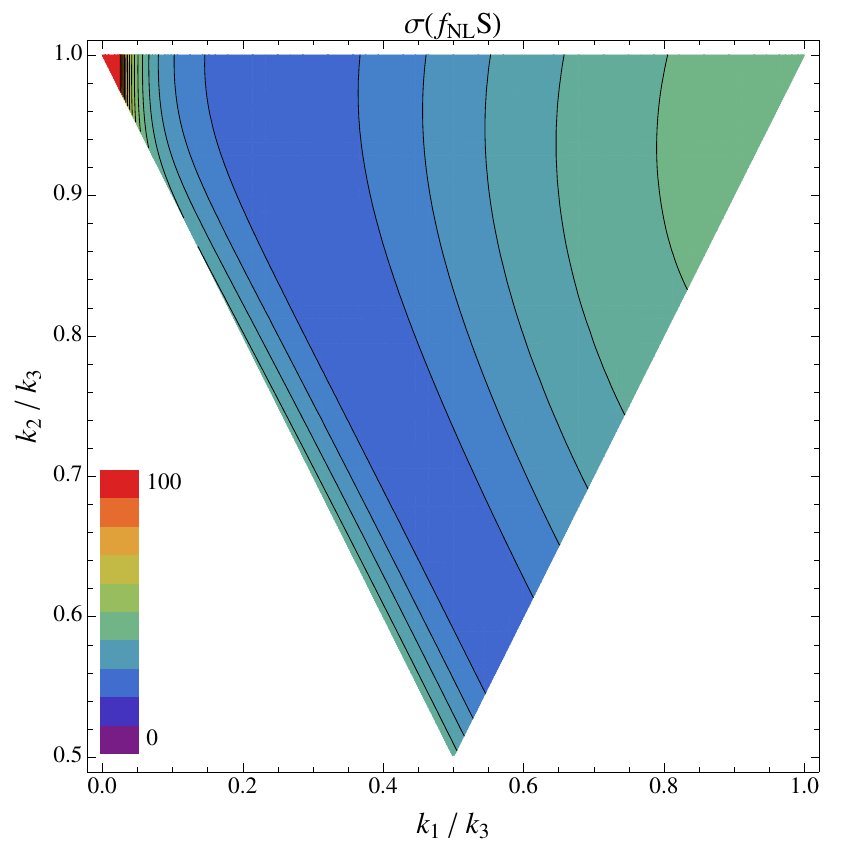}
        }
        \caption{Example contour slices of $\sigma(f_{NL}S)$  under different divergence constraints, [top left] $S_{[-2,-2]}$, [top right] $S_{[-2,1]}$, and [bottom] $S_{[-1,-1]}$.}
    \label{fig:sigmaB_TE}
\end{figure}

The weak $k_3$ dependence implies that the uncertainties at one $k_3$ reasonably reflect the overall uncertainties if one were to marginalize over $k_3$. Figure \ref{fig:sigmaB_TE} shows the error on the $k_3=0.01 \mbox{ Mpc}^{-1}$ slice for three different divergence cases.  The location of minimum $\sigma(f_{NL}S)$ comes from the sum of the eigenmodes that is weighted by each mode's error, which arises out of the complementarity of the degeneracy directions of the PC's.  We find the  location of the best measured configuration is consistent for the  scenarios that diverge as  $x_{sq}^{-2}$ through to a constant in the squeezed limit for $R=-2$. This location is not situated in any one of the corners of the triangle plot associated with squeezed, equilateral, and flattened configurations. Instead it is somewhat centrally located adjacent to the flattened edge. The best-measured configurations are located at $\frac{k_1}{k_3} \approx 0.32$ and $\frac{k_2}{k_3} \approx 0.80$, with minimum $\sigma(f_{NL}S) \approx 37$. For the vanishing divergence prior, the best-measured configuration approaches the squeezed limit, as we have required the noise to scale as the shapes, which go to zero there.  We also find that for shapes constructed from the local, equilateral, and orthogonal templates ($R=-1$), the best measured location spans a degeneracy direction also along the flattened edge, with minimum $\sigma(f_{NL}S) \approx 20$.  

While the best measured region, in which the error is a minimum, is useful in the absence of knowledge about the theory, the signal-to-noise ratio, for a given underlying model can also help determine a survey's ability to distinguish between shapes. We consider this in the following subsection.

\subsection{Shape determination and distinguishability}
\label{sec:shape determination}

We now turn to discussing a central question of the paper: given a detection of non-Gaussianity using a specific template, what can be confidently inferred about the true underlying shape? We have already considered this from one perspective in section \ref{sec:fisher_results} by considering the uncertainties in ascribing a detection using a template to the template's shape itself. If we allow for the possibility that a detection using a specific template could be detecting the component of another shape allowed by the theoretical prior we are considering, then the errors on shape determination can increase significantly, especially for shapes that do not peak in the squeezed or flattened configurations.

In this section we approach the question of shape distinguishability from a second direction,  considering the range of possible general shapes, under a divergence prior, that could create the detected template signal and fit the bispectrum data within some confidence range. Such analyses have already been considered in the context of specific models, for example, how well we might disentangle a QSFI model (e.g. in \eqref{eq:qsfi_shape}) from $S_{equil}$ or $S_{local}$ as a function of $\nu$ \cite{Sefusatti:2012ye,Norena:2012yi}. Here we extend this approach to a more general shape, and consider what implications a detection with one of the common templates has for general models.  For specificity we consider a subset of general shapes consistent with $S_{[-2,1]}$, 
\bea
S_{gen} &=& (1-\alpha_X-\alpha_Y-\alpha_Z)S_{equil}+\alpha_X S_X+\alpha_Y S_Y+\alpha_Z S_Z,
\eea
where $S_{X,Y,Z}$ can be $\{ S_{ortho}+6K_4-6K_3,S_{local}-2K_5+2K_3,2K_3-K_6\}$.  This is investigating a general set of single-field inflation models from which $S_{ortho(2)}$ in \eqref{eq:S_ortho(2)} and $S_{enf(2)}$ in \eqref{eq:S_enf(2)} are drawn.

How large must a template signal $f_{NL}^T$ be to be confident that the signal is not from a different, more general shape $S_{gen}$?  We set this distinguishable detection threshold to be $\sigma(f_{NL}^T)$, the error on $f_{NL}^T$ for the template, marginalized over $f_{NL}^S$, the amplitude of the general shape.  The marginalized constraint is computed by inverting the 2$\times$2 Fisher matrix for $(f_{NL}^T,f_{NL}^S)$.  Thus we are comparing two shapes, where one is a template, and the other is a general shape, in which $\alpha_X$, $\alpha_Y$, and $\alpha_Z$ parametrize the deviation from $S_{equil}$.

\begin{figure}[t]
\centering
{
\includegraphics[width=0.58\textwidth]{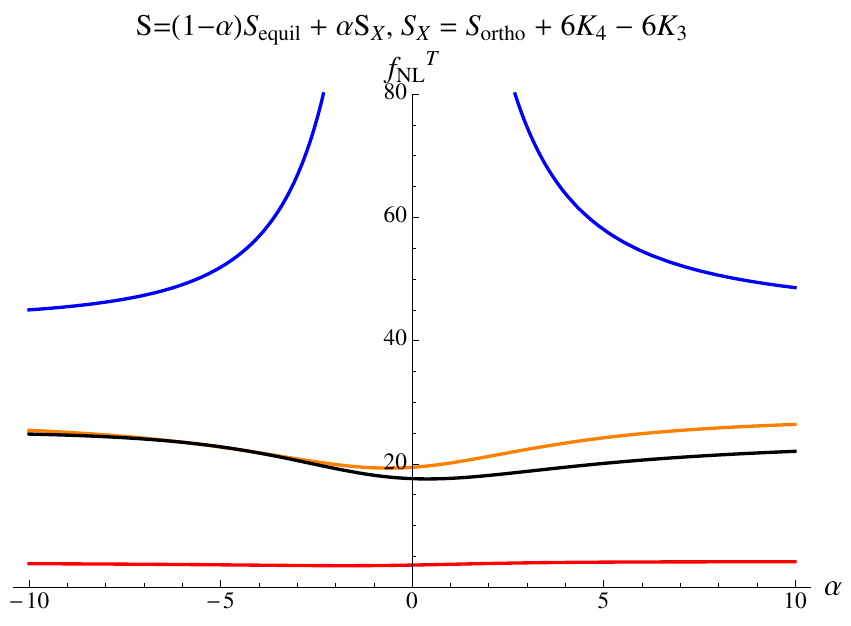}\\
\includegraphics[width=0.58\textwidth]{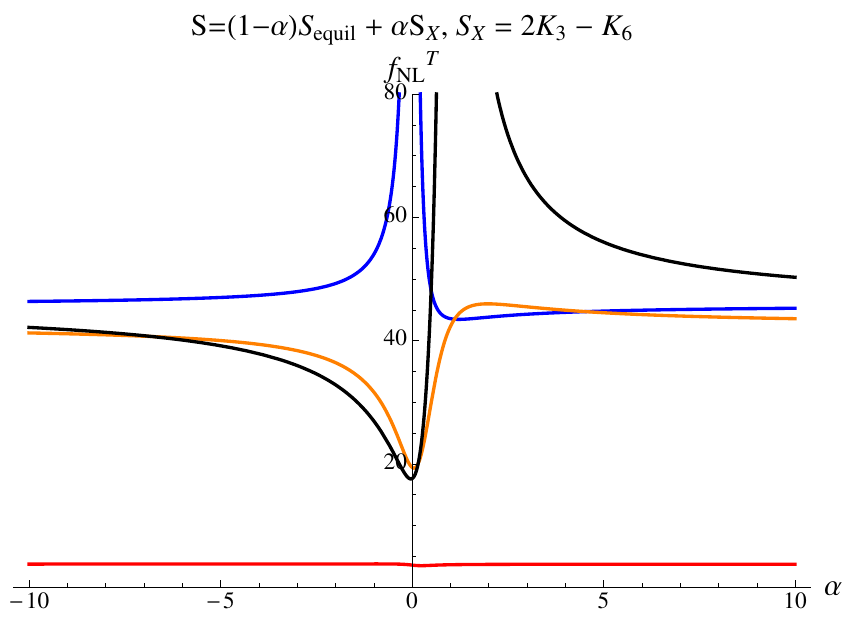}}
       \caption{Detection thresholds on the amplitude of templates, $f_{NL}^T$, for distinguishing between the template and a general shape, $S=(1-\alpha)S_{equil}+\alpha S_{X}$, at the $1\sigma$ confidence level.  [Top] $S_{X} =S_{ortho}+6{\cal K}_4-6{\cal K}_3$, [bottom] $S_{X} =2{\cal K}_3-{\cal K}_6$. Blue, orange, red, and black curves denote $f_{NL}^{equil}$, $f_{NL}^{orth}$, $f_{NL}^{local}$, and $f_{NL}^{2{\cal K}_3-{\cal K}_6}$, respectively.  Since $S_{local}-2{\cal K}_5+2{\cal K}_3$ is very similar to $2{\cal K}_3-{\cal K}_6$, the case where $S_{X} =S_{local}-2{\cal K}_5+2{\cal K}_3$ is not shown.
         \label{fig:template_1D_map}
         } 
\end{figure}

In the simplest case, we allow only $\alpha_X$ to be non-zero, such that $S_{gen}$ is a linear combination of two shapes, $S_{equil}$ and $S_X$, that varies with one parameter.  In Figure \ref{fig:template_1D_map}, we show $\sigma(f_{NL}^T)$ for the  local, equilateral, and orthogonal templates when $S_X$ takes different forms.  The minimum value of $\sigma(f_{NL}^T)$ for each template across all values of $\alpha$ recovers the unmarginalized errors of each.  A detected value of $f_{NL}^{equil}$ must be larger to produce a $1\sigma$ detection of the equilateral shape, as opposed to a more general shape with $\alpha\ne0$, while $f_{NL}^{local}$ never has to be much larger than the unmarginalized $\sigma(f_{NL}^{local})$ to favor the local model over this general shape, because $S_X$ and $S_{local}$ are weakly correlated.  

To illustrate the use of Figure \ref{fig:template_1D_map}, for example, a detection of $f_{NL}^{orth}=40$, while greater than the unmarginalized error of 19, would only be sufficient to rule out a false $1\sigma$ detection of $S_{gen}$ with $S_X = 2{\cal K}_3-{\cal K}_6$ for $-5 \lesssim \alpha \lesssim 0.9$.  On the other hand, if $f_{NL}^{orth}$ is detected to be larger than 46, then $S_{gen}$ of the 1-parameter form would be disfavored, as $\sigma(f_{NL}^{orth})$ is smaller than this over all values of $\alpha$. Models with $S_X = 2{\cal K}_3-{\cal K}_6$ are most easily differentiated from $S_{equil}$ because they have the lowest correlation with $S_{equil}$.

An application of comparing constraints on $\alpha_X$ from two distinct templates is to test whether a given model is consistent with or disfavored by the data. 
 If two template measurements each individually remain consistent with two non-overlapping regions of $\alpha$-space, then it would be clear that modeling the underlying shape with $\alpha$ alone is not able to produce a viable model.  
This would be true for dual measurements of $\{f_{NL}^{equil}=60,f_{NL}^{ortho}=45\}$ for $S_X=2{\cal K}_3-{\cal K}_6$, since they would imply non-overlapping ranges of $\alpha$, $-0.7 \leq \alpha \leq 0.3$ versus $1.3 \leq \alpha \leq 3.8$ to each be consistent with the data. 

\begin{figure}[t]
\centering
\includegraphics[width=0.4\textwidth]{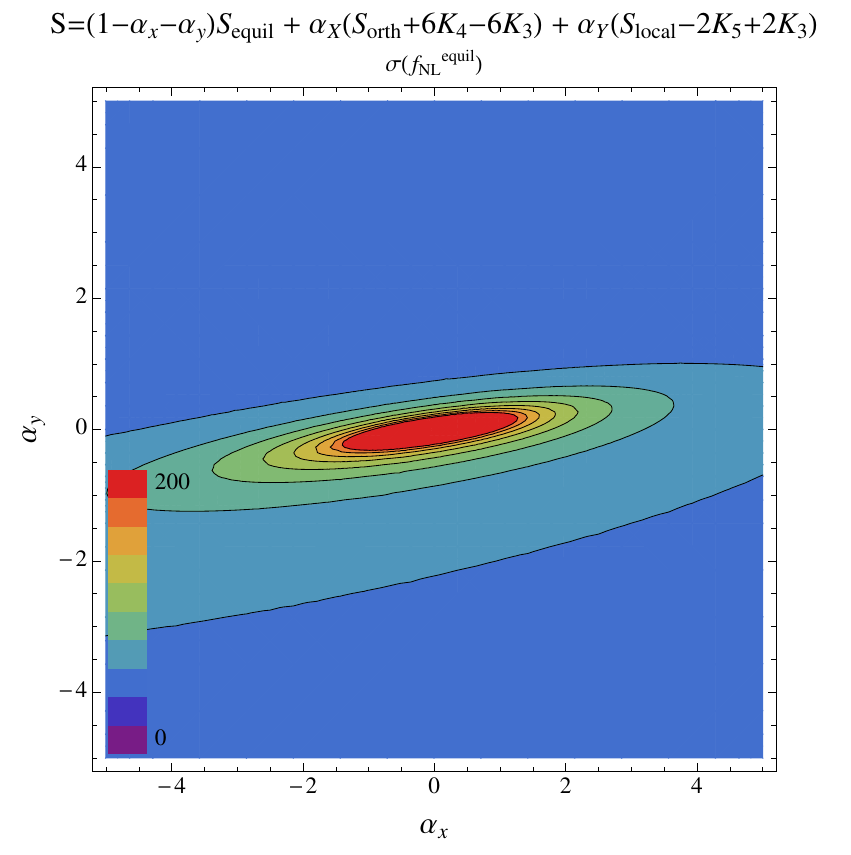}
\includegraphics[width=0.4\textwidth]{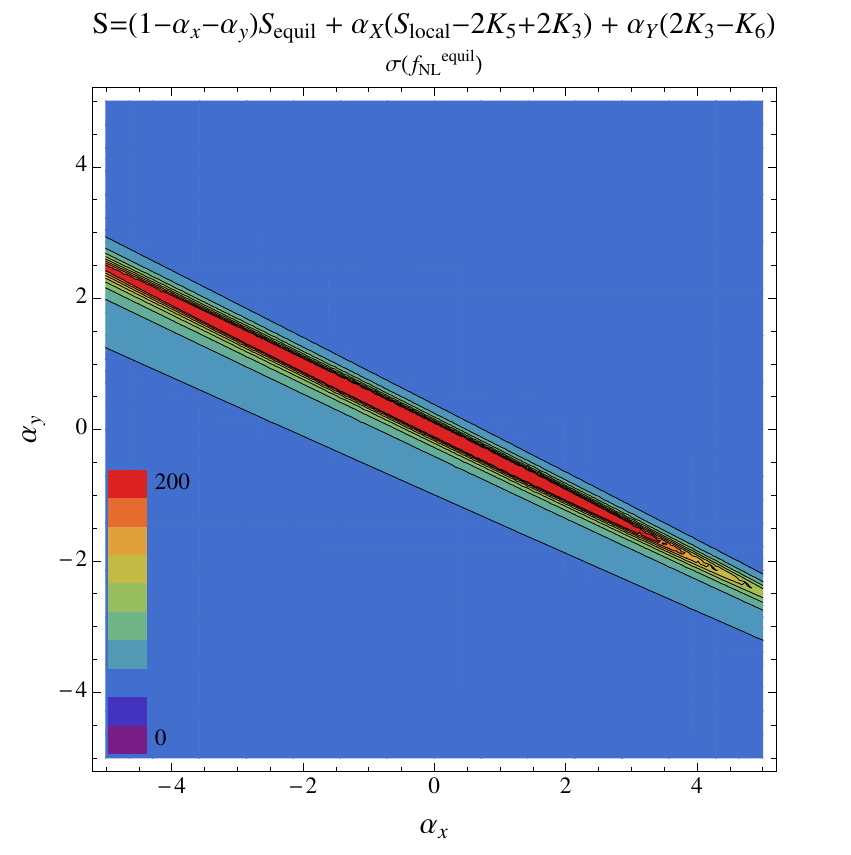}
\includegraphics[width=0.4\textwidth]{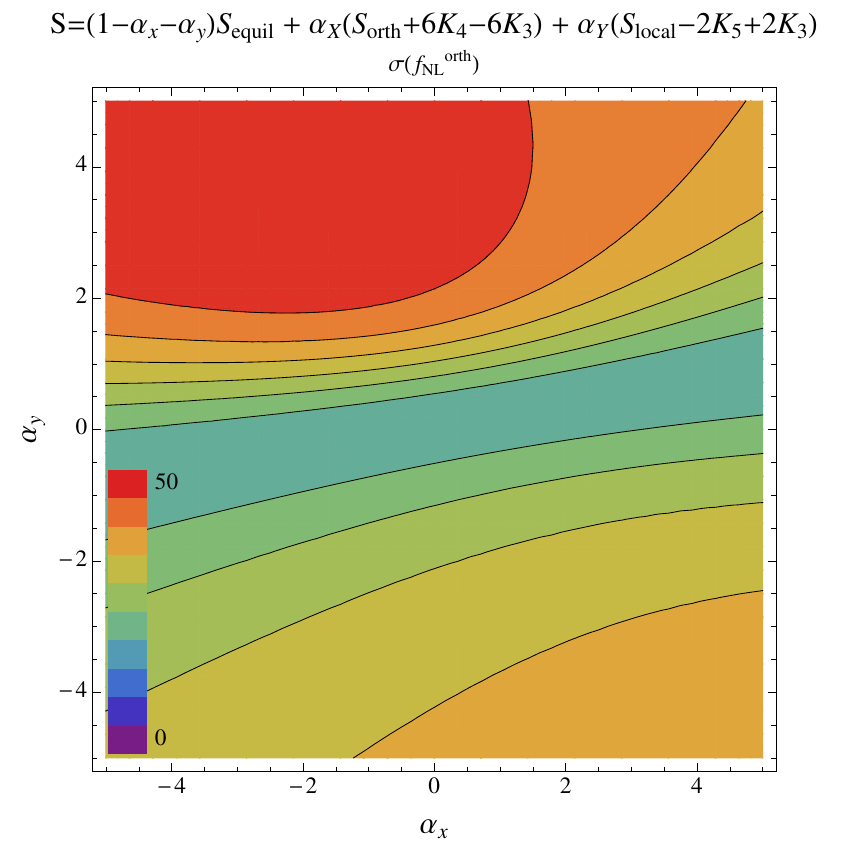}
\includegraphics[width=0.4\textwidth]{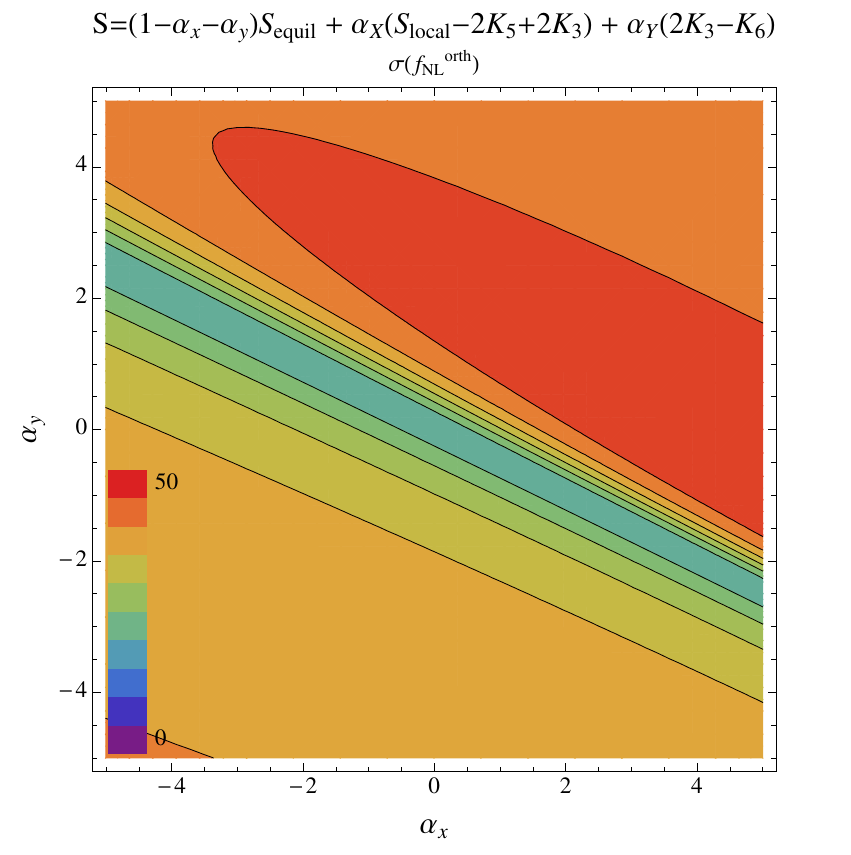}
\includegraphics[width=0.4\textwidth]{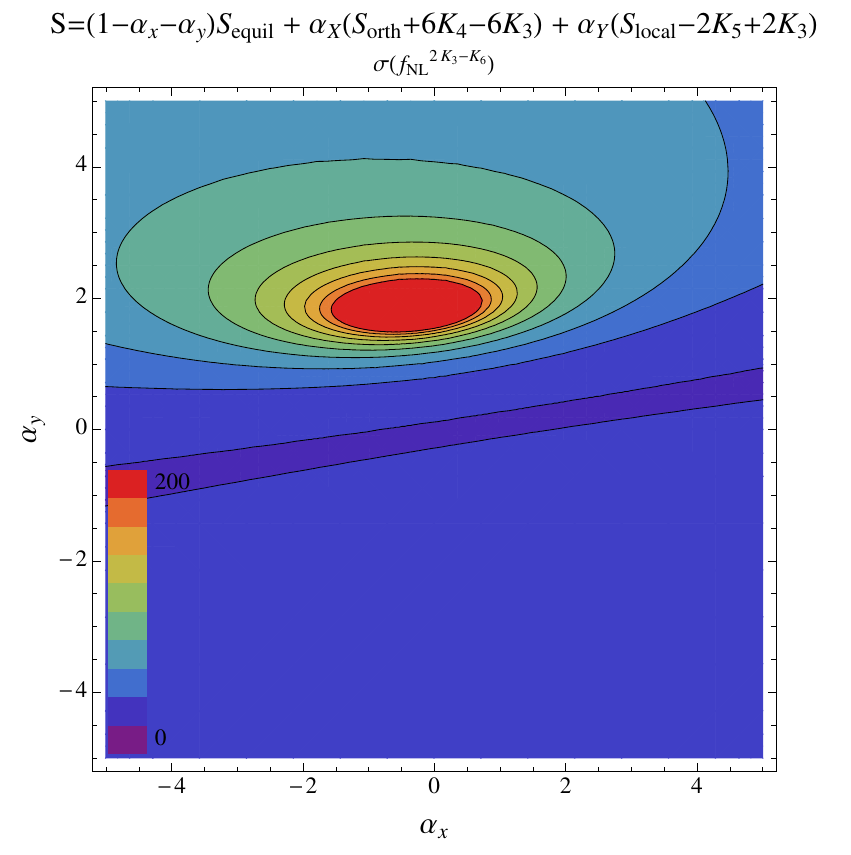}
\includegraphics[width=0.4\textwidth]{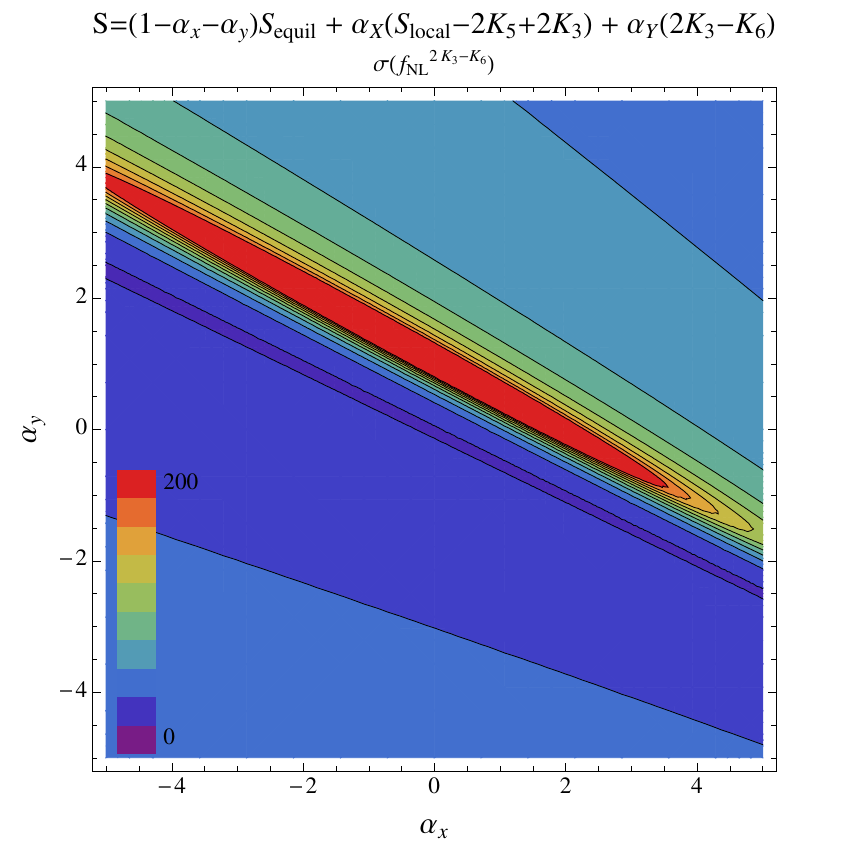}
        \caption{  Detection thresholds on the amplitude of templates, [top] $f_{NL}^{equil}$, [center] $f_{NL}^{orth}$, and [bottom] $f_{NL}^{2{\cal K}_3-{\cal K}_6}$, for distinguishing between each template and two forms of a general shape $S$ at the $1\sigma$ confidence level.  The general shapes considered are [left panels] $S_{gen} = (1-\alpha_X-\alpha_Y)S_{equil}+\alpha_X (S_{ortho}+6{\cal K}_4-6{\cal K}_3)+\alpha_Y (S_{local}-2{\cal K}_5+2{\cal K}_3)$ or [right panels] $S_{gen} = (1-\alpha_X-\alpha_Y)S_{equil}+\alpha_X (S_{local}-2{\cal K}_5+2{\cal K}_3)+\alpha_Y (2{\cal K}_3-{\cal K}_6)$ .  Contours for $f_{NL}^{loc}$ are not pictured, because the marginalized $\sigma(f_{NL}^{loc})$ remains close to its unmarginalized value over these 2-dimensional spaces.  The case where $S_X = S_{ortho}+6{\cal K}_4-6{\cal K}_3$ and $S_Y = 2{\cal K}_3-{\cal K}_6$ is not pictured because $S_X$  and $S_Y$ are nearly uncorrelated, thus no additional information is revealed beyond that in Figure \ref{fig:template_1D_map}.
        \label{fig:template_2D_map} } 
\end{figure}

We can extend the same analysis to a comparison between templates and a 2-parameter general shape by allowing both $\alpha_X$ and $\alpha_Y$ to vary simultaneously, while $\alpha_Z$ is fixed to zero.  For example, $S_{enf(2)}$ is a specific template for which this is true.  In Figure \ref{fig:template_2D_map} we show $\sigma(f_{NL}^{equil})$ and $\sigma(f_{NL}^{orth})$ over different choices of the 2-dimensional space and find that there exist degeneracy directions that are not fully captured by the 1-dimensional projections in Figure \ref{fig:template_1D_map}.  We find that $\sigma(f_{NL}^{local})$ remains close to the unmarginalized value in this case as well.

In the most general 3-parameter model, we can ask the question of whether there is any area of this space corresponding to a general model that vanishes in the squeezed limit, with a significant enough overlap with the local template to require that a potentially detected $f_{NL}^{local}$ be much greater than the unmarginalized value of 3.  If this were the case, then it may be that a local template detection cannot definitively rule out a general shape that satisfies the single-field consistency relation.  However, we find that nowhere in the parameter space does the $\sigma(f_{NL}^{loc})$ become greater than 4.2, showing that a detection of the local template above this threshold would effectively rule out a general shape, vanishing in the squeezed limit, subject to the assumption that it can be written in terms of our basis in $S_{[-2,1]}$.  The same distinguishing power is not present for $S_{local}$ if we allow a weaker prior given by $S_{[-2,-1]}$. In this case the significant cosine between $S_{local}$ and $2{\cal K}_5-{\cal K}_6$, means we may never be able to confidently attribute a detection with $S_{local}$ to be definitive evidence that the diverging signal is unambiguously $S_{local}$.  A long shot could be to additionally look at the correlation of the bispectrum signal with $2{\cal K}_4-{\cal K}_6$ which is mildly negatively correlated with $S_{local}$ and essentially uncorrelated with $2{\cal K}_5-{\cal K}_6 $.

This last point raises an interesting application of our study: to ask if there are distinct, new templates that we might use to learn about the origins of a detected non-Gaussian signal. In the context of models described by the first three modes, ${\cal K}_0$ to ${\cal K}_2$, the local, equilateral, and orthogonal templates are almost perfectly aligned with the principal components. If we extend the templates to include  ${\cal K}_3$ through ${\cal K}_6$, however, we find these no longer represent the PC's. For example, what might be the best way to  extend the template pool to search for signatures of single-field inflation models with Bunch-Davies vacua? In the context of $r=1$ shapes, $2{\cal K}_3-{\cal K}_6$ is well-aligned with the best measured PC and is only mildly correlated with the existing templates which would make it a reasonable candidate to add as an additional template. We show the resulting constraints on general shapes in Figures \ref{fig:template_1D_map} and \ref{fig:template_2D_map}.  The figures show that this template probes regions of the allowed $\alpha$-space which the equilateral and orthogonal templates do not constrain in the same way.  Thus it may be possible to combine constraints from the common templates and motivated choices of a small number of new templates, like $2{\cal K}_3-{\cal K}_6$, to probe the underlying shape of non-Gaussianity. 


\section{Conclusion}
\label{conclusion}

At the heart of this work is the discussion about how uncertainties quoted on shape detection are inherently dependent on the underlying assumptions made about the shape. While a detection of non-Gaussianity with any template will be extraordinarily transformative in our field, its interpretation, in what it tells us about the underlying shape, has to be considered carefully in terms of our underlying theoretical prior we impose. Even if no detection of non-Gaussianity is made, upper bounds on the deviations from Gaussianity according to templates will have broader impacts for constraints on general shapes.

We have presented an approach for quantifying how well upcoming CMB temperature and $E$-mode polarization data can determine the shape of primordial non-Gaussianity under minimal assumptions.  
We proposed a set of polynomial divergent basis functions, $\{K_n\}$, that are well-tuned to describing many nearly scale-invariant, smoothly varying, but potentially divergent shapes discussed in the literature. 
We find we need only three to seven modes to generate matched templates to describe a wide range of physically motivated shapes. In this sense, the divergent basis is more efficient than the polynomial basis used in previous studies (e.g. \cite{Fergusson:2009nv}).
Each $\mathcal{K}_n$ in our basis is generally divergent, but linear combinations of the $\mathcal{K}_n$ can be constructed to have cancellations in the squeezed limit, thus creating templates that are less divergent (e.g. equilateral shape).   For example, $S_{equil}$ and $(2\mathcal{K}_3-\mathcal{K}_6)$ both vanish in the squeezed limit, but still have a low correlation because the latter has more power near the flattened and squeezed configurations.

Using the $\{\mathcal{K}_n\}$ it is straightforward to form template classes, $S_{[R,r]}$, that have specific, common divergence properties in the squeezed limit. Each class is constructed from an irreducible set of shapes, that while constructed out of a basis sets  with maximum divergence $x_{sq}^{R}$ ($R<0$), through cancellations of divergent terms, have squeezed limit  $x_{sq}^r$ $(r>R)$. The choice of $R$ controls how many basis modes are used to develop the templates, e.g. $R=-1$ includes ${\cal K}_{0}$ through ${\cal K}_{2}$, while $R=-2$ uses ${\cal K}_{0}$ through ${\cal K}_{6}$. As $R$ becomes more negative it allows  templates to be refined and shapes with a broader set of features across configuration space to be modeled.  

The classes allow templates to be developed with priors that are well-motivated by theories: $S_{[R,1]}$ represents the class of all single-field models derived from a Bunch-Davies vacuum, $S_{[R,-1]}$ in addition includes all multi-field models that diverge like the local shape in the squeezed limit, and $S_{[R,-2]}$ is the most general class of shapes which includes models from non-Bunch-Davies vacuum initial states.  

While the constituent shapes making up each class have the same divergence properties in the squeezed limit, away from this limit they have  power weighted differently in the configuration space. For example, we discuss a new shape, $2{\cal K}_{3}-{\cal K}_{6}$, used in $S_{[R,1]}$, that has the same squeezed limit behavior as the equilateral shape but has $\ell$-space cosines with the standard equilateral, orthogonal, and local templates of 0.07, 0.80 and -0.29 respectively. While the divergent terms cancel in the squeezed limit, $2{\cal K}_{3}-{\cal K}_{6}$  has significant power just away from the squeezed and flattened configurations that differentiates it from the equilateral shape, and leads to it being most similar (though only mildly) with the orthogonal template.

An added benefit of using the divergent basis and template classes to consider general shapes  is that it ties together the methods we use to search for evidence of shapes with CMB data to LSS constraints from a scale-dependent halo bias, which probes the squeezed limits of shapes.  It is well-known that templates for physical shapes which work for generating CMB predictions can fail when used for LSS predictions \cite{Wagner:2011wx}, because while CMB constraints represent a weighted average over all $k$-space configurations, the halo bias traces the squeezed limit region of $k$-space only.  Thus our approach provides a way of generating templates that can potentially be used consistently for both CMB and LSS studies.

We adopt a Fisher matrix approach modeled on a Planck-like survey to estimate uncertainties on the amplitudes of shapes within each shape class, $r$.  As summarized in Table \ref{tab:sigmafNL}, we computed the uncertainties on shape attribution under each prior and how these uncertainties on confidently being able to determine that a template is the true shape can change substantially dependent upon the type of prior we impose. 

We find that the best measured shapes are those with the strongest divergence and with principal power near squeezed and flattened $k$-configurations.  Though the conventional approach is to quote constraints at the equilateral configuration, $k_1=k_2=k_3$, we show,  as summarized in Table \ref{table:pca_te}, that this convention can mask how well or badly a shape is measured, as doing so has the effect of re-normalizing constraints such that badly measured modes can appear to have constraints similar to the best measured mode.

Using the PCA results, we map out the $k$-dependence of the constraints for a general shape given a prior, and show its dependence on the prior.  For all but the $r=1$ case, the best measured location is not in the equilateral configuration where shapes and constraints on $f_{NL}$ are typically normalized, but in a configuration that is neither squeezed, flattened, or equilateral, but somewhere in between.  This best measured location at roughly $k_1/k_3\approx0.32$ and $k_2/k_3\approx0.80$ arises out of the complementary gradients of the power in the PC's.  For the $r=1$ case, the best measured location is weighted more strongly towards the squeezed configuration, reflecting that the signal and the noise, with which it is correlated, both go to zero in this limit.

Given our parametrization of a general shape under a divergence prior, we then ask how well it could be constrained using measurements of amplitudes of common templates, like the local, equilateral, and orthogonal templates.  We focus on the class of general shapes that can represent the possible range of single-field models that vanish in the squeezed limit ($r=1$).  We calculate bounds on the subset of shapes that can remain consistent with constraints on the local, equilateral, and orthogonal templates, and find again--consistent with what we found earlier in the analysis--that templates with more power in the squeezed and flattened configurations provide more stringent constraints on this class of general shapes.  Thus, the local, equilateral, and orthogonal templates serve different roles in constraining general shapes; the local template, if detected with sufficient amplitude, will rule out any shape of this type, while the equilateral and orthogonal templates serve to put constraints around different regions of the parameter space.  In this sense, constraints from different templates can be complementary.

Furthermore, a general (unknown) shape, will have different overlaps with the templates, creating a possibility that by combining constraints on templates, the overall constraint will shed more light on the underlying theory than any one constraint alone. We find it can also be advantageous to look for signals with a new, distinct template, beyond the three standard ones, that could help constrain models more efficiently; we explored the potential for using $2{\cal K}_3-{\cal K}_6$ in this context.

In this initial study we use somewhat idealized assumptions focusing on the effects of cosmic variance and Gaussian noise from a homogeneous sky coverage. We recognize the rich potential for further study to other basis sets, that better characterize sharp or oscillatory features in bispectra, the presence of isocurvature modes, and stronger deviations from scale-invariance. To confidently attribute a  primordial source to any measured non-Gaussianity one would also want to fully account for contributions from  astrophysical and instrumental sources, including gravitational lensing, inhomogeneous sky coverage, and secondary anisotropies from astrophysical foregrounds. There is also the substantial question of how large-scale structure measurements, with sensitivity to the squeezed limit, can complement the CMB data in constraining these general shapes, as well as whether 4-point statistics and checks of non-Gaussian consistency ansatzes can play a role. We are tackling some of these intriguing issues in work in preparation. 

\section{Acknowledgements}

We would like to thank Nishant Agarwal, Xingang Chen, Tom Loredo, Liam McAllister, Sarah Shandera, and the anonymous referee for useful discussions during the preparation of this paper. JB and RB's research was supported by NSF CAREER grant AST0844825, NASA Astrophysics Theory Program grants NNX08AH27G and NNX11AI95G and by Research Corporation. 


\bibliographystyle{JHEP}
\bibliography{NGrefs}

\end{document}